\documentclass[reprint,superscriptaddress,
amsmath,amssymb,
aps,
prl,
]{revtex4-2}

\usepackage{graphicx}
\usepackage{dcolumn}
\usepackage{bm}
\usepackage{natbib}
\usepackage[]{lineno}

\usepackage[backref=none,bookmarksnumbered=true,bookmarks=true,bookmarksopen=true,colorlinks=true,
citecolor=blue,linkcolor=blue,anchorcolor=green,urlcolor=blue,unicode=false]{hyperref}
\usepackage{ulem}[normalem] 
\usepackage{siunitx}
\usepackage{titlesec}
\usepackage{setspace}

\newcommand{\dd}{\mathop{}\!\mathrm{d}}

\newcommand{\rev}[1]{\textcolor{black}{ #1}}
\def\0t{\textbf{AT}}
\def\1t{\textbf{AAT}}
\def\2t{\textbf{BAT}}
\def\3t{\textbf{TAT}}

\begin{document}

\title{Topological Engineering of a Frustrated Antiferromagnetic Triradical in Aza-Triangulene Architectures}
\author{Francisco Romero-Lara}\thanks{These authors contributed equally to this work.}
    \affiliation{CIC nanoGUNE-BRTA, 20018 Donostia-San Sebasti\'an, Spain}
\author{Manuel Vilas-Varela}\thanks{These authors contributed equally to this work.}
    \affiliation{Centro Singular de Investigaci\'on en Qu\'imica Biol\'oxica e Materiais Moleculares (CiQUS) and Departamento de Qu\'imica Org\'anica, Universidad de Santiago de Compostela, 15782 Santiago de Compostela, Spain}
\author{Manish Kumar}\thanks{These authors contributed equally to this work.}
    \affiliation{Institute of Physics of the Czech Academy of Sciences, Cukrovarnicka 10, Prague 6, CZ 16200, Czech Republic}
\author{Ricardo Ortiz}\thanks{These authors contributed equally to this work.}
    \affiliation{Donostia International Physics Center (DIPC), 20018 Donostia-San Sebastian, Spain}
    \affiliation{CICECO- Aveiro Institute of Materials, Department of Chemistry, University of Aveiro, 3810-193 Aveiro, Portugal}
\author{Alessio Vegliante}
    \affiliation{CIC nanoGUNE-BRTA, 20018 Donostia-San Sebasti\'an, Spain}
\author{Lucía Gómez-Rodrigo}
    \affiliation{Centro Singular de Investigaci\'on en Qu\'imica Biol\'oxica e Materiais Moleculares (CiQUS) and Departamento de Qu\'imica Org\'anica, Universidad de Santiago de Compostela, 15782 Santiago de Compostela, Spain}
\author{Trisha Sai}
    \affiliation{CIC nanoGUNE-BRTA, 20018 Donostia-San Sebasti\'an, Spain}
\author{Jan Patrick Calupitan}
    \affiliation{Donostia International Physics Center (DIPC), 20018 Donostia-San Sebastian, Spain}
    \affiliation{Centro de F\'isica de Materiales (CFM-MPC), Centro Mixto CSIC-UPV/EHU, E-20018 Donostia-San Sebasti\'an,  Spain}
    \affiliation{Sorbonne Université, CNRS, Institut Parisien de Chimie Moléculaire (IPCM), F-75005 Paris, France}
\author{Diego Soler}
    \affiliation{Institute of Physics of the Czech Academy of Sciences, Cukrovarnicka 10, Prague 6, CZ 16200, Czech Republic}
\author{Niklas Friedrich}
    \affiliation{CIC nanoGUNE-BRTA, 20018 Donostia-San Sebasti\'an, Spain}
\author{Dongfei Wang}
    \affiliation{CIC nanoGUNE-BRTA, 20018 Donostia-San Sebasti\'an, Spain}
    \affiliation{Key Laboratory of Advanced Optoelectronic Quantum Architecture and Measurement, School of Physics, Beijing Institute of Technology, 100081 Beijing, China}
\author{Jon Ortuzar}
    \affiliation{CIC nanoGUNE-BRTA, 20018 Donostia-San Sebasti\'an, Spain}
\author{Leonard Edens}
    \affiliation{CIC nanoGUNE-BRTA, 20018 Donostia-San Sebasti\'an, Spain}
\author{Stefano Trivini}
    \affiliation{CIC nanoGUNE-BRTA, 20018 Donostia-San Sebasti\'an, Spain}
\author{Thomas Frederiksen}
    \affiliation{Donostia International Physics Center (DIPC), 20018 Donostia-San Sebastian, Spain}
    \affiliation{Ikerbasque, Basque Foundation for Science, 48009 Bilbao, Spain}
\author{Fabian Schulz}
    \affiliation{CIC nanoGUNE-BRTA, 20018 Donostia-San Sebasti\'an, Spain}
\author{Pavel Jelínek}\email{jelinekp@fzu.cz}
    \affiliation{Institute of Physics of the Czech Academy of Sciences, Cukrovarnicka 10, Prague 6, CZ 16200, Czech Republic}
\author{Diego Peña}\email{diego.pena@usc.es}
    \affiliation{Centro Singular de Investigaci\'on en Qu\'imica Biol\'oxica e Materiais Moleculares (CiQUS) and Departamento de Qu\'imica Org\'anica, Universidad de Santiago de Compostela, 15782 Santiago de Compostela, Spain}
    \affiliation{Oportunius, Galician Innovation Agency (GAIN), 15702 Santiago de Compostela, Spain}
\author{José Ignacio Pascual}\email{ji.pascual@nanogune.eu}
    \affiliation{CIC nanoGUNE-BRTA, 20018 Donostia-San Sebasti\'an, Spain}
    \affiliation{Ikerbasque, Basque Foundation for Science, 48009 Bilbao, Spain}

\date{\today}

\begin{abstract}  
\textbf{Open-shell nanographenes provide a versatile platform to host unconventional magnetic states within their $\pi$-conjugated networks. Particularly appealing are graphene architectures that incorporate spatially separated radicals and tunable interactions, offering a scalable route toward spin-based quantum architectures. Triangulenes are ideal for this purpose, as their radical count scales with size, although strong hybridization prevents individual spin control. Here, we realize a radical reconfiguration strategy that transforms a single-radical aza-triangulene into three topologically protected spin states by covalently extending it with armchair anthene moieties of increasing length. \rev{Scanning tunnelling spectroscopy reveals emergent correlated spins forming a frustrated spin trimer whose interaction weakens with anthene length. This trend is captured by multi-reference electronic-structure calculations, which trace a progressive rise in polyradical character driven by the progressive reorganization of the correlated frontier orbitals into three edge-localized natural orbitals. Consequently, the initial single-radical doublet reorganizes  into  non-interacting edge spins,} a molecular analog of a three-qubit quantum register.}
\end{abstract}

\maketitle



Open-shell nanographenes provide a chemically precise route to engineer spin arrays for quantum information technologies. Their $\pi$-conjugated lattices can host spatially separated electronic spins that are naturally protected from decoherence due to weak spin–orbit and hyperfine couplings \cite{YAZYEV_Emergence_2010,OTEYZA_Carbonbased_2022,YAO_Spinorbit_2007,YAZYEV_Magnetic_2008,YAZYEV_Hyperfine_2008}. The possibility of encoding and coherently manipulating such spin states makes nanographenes attractive building blocks for molecular quantum registers \cite{WANG_Topological_2009}. 

Triangulene represents one of the most prototypical open-shell polycyclic aromatic hydrocarbons (PAHs) due to its non-Kekulé structure and a spin ground state that correlates directly with its size \cite{CLAR_Aromatic_1953}.
Recent advances in on-surface synthesis \cite{CLAIR_Controlling_2019} have enabled the successful fabrication of triangulenes \cite{PAVLICEK_Synthesis_2017, MISHRA_Synthesis_2019, SU_Atomically_2019, MISHRA_Synthesis_2021, TURCO_Observation_2023}, targeting spin-state control through molecular size and the design of covalently linked architectures \cite{MISHRA_Collective_2020,MISHRA_Observation_2021,HIEULLE_OnSurface_2021,CHENG_Onsurface_2022,KRANE_Exchange_2023,YUAN_Fractional_2025,SU_Fabrication_2025,ZHAO_Spin_2025,VEGLIANTE_OnSurface_2025,TURCO_Multiconfigurational_2025}. In triangulenes, all radical centers reside on the same sublattice, leading to strong wavefunction overlap and ferromagnetic-like exchange interactions that maximize the total spin, in accordance with Lieb's theorem~\cite{LIEB_Two_1989,FERNANDEZ-ROSSIER_Magnetism_2007}.  

\begin{figure*}[t!]
        \centering
    	\includegraphics[width=17.4cm]{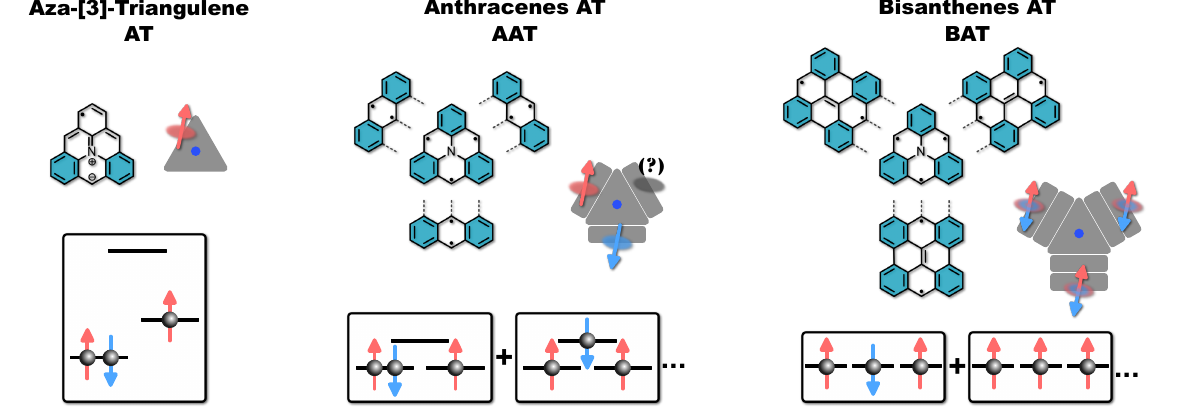}
        \caption{\label{fig:Fig1} \textbf{Construction of the extended aza-triangulene platforms through anthene addition.} Progression of open-shell states in aza-[3]-triangulene architectures by incorporation of anthene units with increasing length. The aromatic Clar sextets are colored in light blue. The structures evolve from a monoradical, to a frustrated antiferromagnetic trimer, to an independent topological triradical.  A schematic representation of active energy states and their electron/spin redistribution found in this work is added at the bottom. The extensions afford an increased correlation of the spin states}
 \end{figure*}

Introducing antiferromagnetic exchange pathways between conjugated spins opens new opportunities to induce frustration within the $\pi$-network~\cite{MISHRA_Topological_2020,SONG_Highly_2024,ZHAO_Tunable_2024,LI_Designer_2025}, thereby favoring entangled spin states in the ground state. A particularly effective strategy to achieve this involves substitutional doping, which modifies the sublattice symmetry, the spin ground state, and their exchange interactions without disrupting the conjugated framework of the nanographene~\cite{WANG_AzaTriangulene_2022,VILAS-VARELA_OnSurface_2023,LAWRENCE_Topological_2023,CALUPITAN_Emergence_2023,LI_OnSurface_2024,HENRIQUES_Spin_2024}. 
For instance, replacing the central carbon of a [3]-triangulene with nitrogen,  i.e. in aza-[3]-triangulene (\0t), introduces an additional $\pi$-electron into the majority-spin manifold. While this configuration might naively be expected to stabilize a high-spin triradical state, in practice the extra electron occupies the pair of degenerate zero-energy states (ZES) of the [3]-triangulene, resulting in a single radical and spin-doublet ground state (Fig.~\ref{fig:Fig1}) that is further stabilized by a Jahn–Teller distortion~\cite{SANDOVAL-SALINAS_Triangular_2019,WANG_AzaTriangulene_2022}.

Here, we present a strategy for transforming the spin-doublet \0t molecule \cite{WANG_AzaTriangulene_2022,SANDOVAL-SALINAS_Triangular_2019} into a triradical platform 
by covalently extending its zigzag edges with anthracenyl fragments of length $n$, as illustrated in Fig.~\ref{fig:Fig1}. The resulting architectures mimic a junction of three 7-carbon-wide armchair graphene nanoribbons (7AGNRs) connected through the \0t core. \rev{It has been shown that 7AGNRs with $n>$2 are prone to develop radical states localized at their zigzag termini \cite{WANG_Giant_2016}, originated from the non-trivial topology of the conjugated network \cite{CAO_Topological_2017} and stabilized by the increasing number of Clar sextets with the ribbon's length \cite{CLAR_Aromatic_1983,AJAYAKUMAR_pExtended_2022,TRINQUIER_Predicting_2018}}. Connecting the 7AGNRs through the \0t core reorganizes the energy balance and electron occupation of the ZESs, thereby favoring a triradical ground state. 

We demonstrate this polyradical progression experimentally by synthesizing and studying the \1t ($n=1$) and \2t ($n=2$) nanographenes, shown in Fig.~\ref{fig:Fig1}. We combine low-temperature scanning tunneling microscopy (STM) and multi-reference quantum chemical calculations to confirm that these two molecules are located right at the transition \rev{from a frustrated trimer to three non-interacting spin-½ states. The presence of spin frustration despite the three radicals residing on the same sublattice is attributed to nitrogen doping, which} critically reverses the exchange pattern, leading to weak antiferromagnetic coupling. 
The presented strategy of merging chemical design with topological transitions \rev{represents a chemically controlled pathway to build up molecular platforms with a correlated manifold of interacting spins and tunable interactions, with potential as a quantum register.}

  \begin{figure*}[th!!]
        \centering
    	\includegraphics[width=17.4cm]{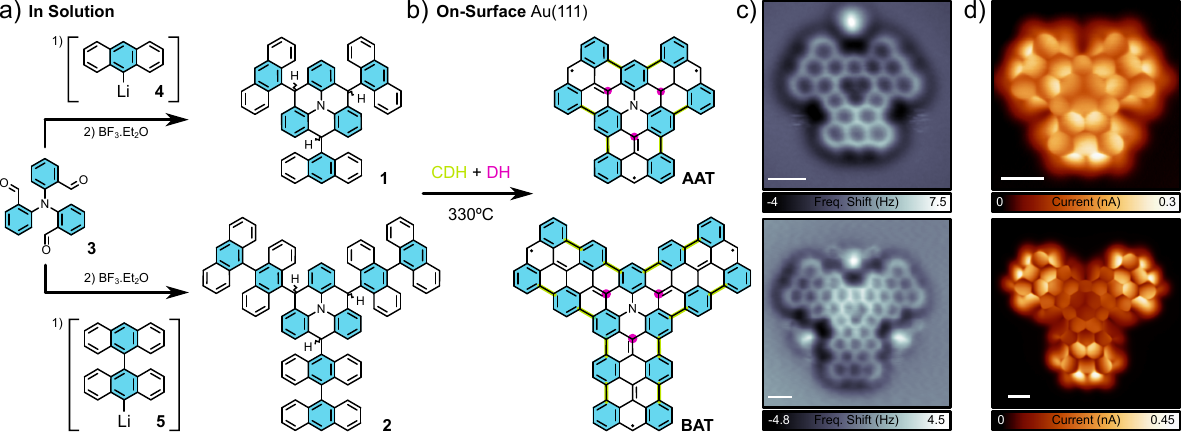}
        \caption{\label{fig:Fig2} \textbf{Synthesis route of extended AT platforms.} a) Chemical reactions performed in solution to yield the targeted precursors \textbf{1} (top) and \textbf{2} (bottom). b) Reactions on-surface of cyclodehydrogenation (CDH, \rev{highlighted in green}) and dehydrogenation (DH, \rev{highlighted in pink}) leading to the desired planar nanostructures. c)  nc-AFM and d) STM images of extended aza platforms \1t (top) and \2t (bottom) performed with a CO-functionalized tip at constant height with $V=0\,\mathrm{mV}$ and $2\,\mathrm{mV}$, respectively (scalebars $4\,\text{\si{\angstrom}}$). The additional bright features in the bay regions of the molecules visible in the nc-AFM images are CO molecules adsorbed next to the nanographenes.}
 \end{figure*}
 
\section*{\label{results_discussion}\raggedright RESULTS}\vspace{-0.5cm}
\noindent\textbf{Synthesis of AT-anthene platforms}\\
The synthesis of \1t and \2t was achieved via the on-surface reaction of the \0t derivatives \textbf{1} and \textbf{2} shown in Fig.~\ref{fig:Fig2}a.
The synthetic strategy followed to obtain these precursors is inspired by the preparation of aza-[5]-triangulene~\cite{VILAS-VARELA_OnSurface_2023}, \rev{and detailed in Supplementary Section S1}. The synthesis involved a sequence of solution-phase reactions, starting with the treatment of tribenzaldehyde \textbf{3} with an excess of organolithium reagents \textbf{4} or \textbf{5}. Then, BF$_3$-promoted, threefold intramolecular Friedel–Crafts cyclization, led to the formation of the \0t derivatives \textbf{1} or \textbf{2}, respectively.

Precursors \textbf{1} and \textbf{2} were deposited on an atomically clean Au(111) surface held at room temperature by sublimation from a silicon wafer (see Methods). Subsequent annealing at $330\,^\circ$C induced substrate-catalyzed dehydrogenation and cyclodehydrogenation reactions~\cite{CLAIR_Controlling_2019}, yielding the targeted aza-[3]-triangulene derivatives shown in Fig.~\ref{fig:Fig2}b. 
The formation of \1t and \2t was confirmed by bond-resolved (BR) non-contact atomic force microscopy (nc-AFM) \cite{GROSS_Chemical_2009} and STM using a CO-functionalized tip (details in Methods). The nc-AFM images (Fig.~\ref{fig:Fig2}c) reveal the honeycomb lattice of the carbon backbone, with the central nitrogen atom appearing darker, consistent with previous reports on aza-[5]triangulene  \cite{VILAS-VARELA_OnSurface_2023, LAWRENCE_Topological_2023}. \rev{Contrary to \0t molecules on gold \cite{WANG_AzaTriangulene_2022}, \1t and \2t remain neutral on the Au(111) surface (see Supplementary Section S5). Sublimation also produced molecular fragments, studied in sections S2 and S3.}

Bond-resolved current images in Fig.~\ref{fig:Fig2}d display 
enhanced signal surrounding the molecular zigzag edges of both \1t and \2t architectures. This local enhancement of the tunneling current indicates an increase of the density of states near the Fermi energy at the zigzag termini. As shown below, this feature originates from the intricate spin structure of the poly-anthracene–extended aza-triangulene core.\\

\begin{figure*}[th!]
        \centering
   	\includegraphics[width=17.4cm]{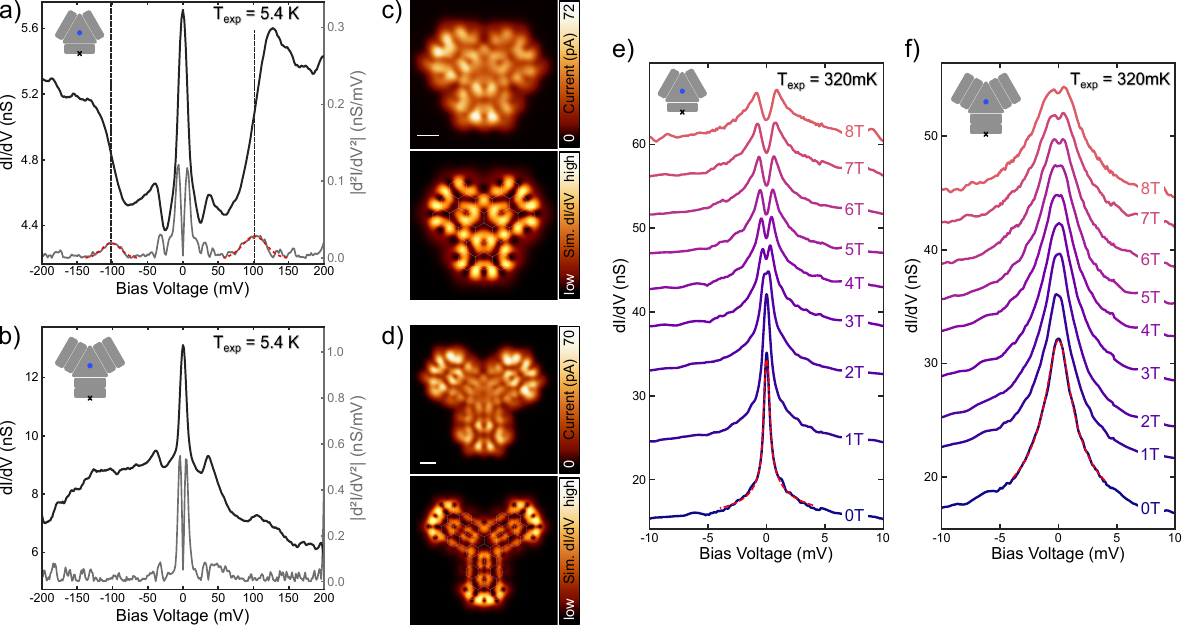}
        \caption{\label{fig:Fig3} \textbf{Detection of $\pi$-magnetism.} a,b) Low-energy $\dd I/\dd V$ spectra measured over a zigzag edge of \1t and \2t, respectively, with a CO-functionalized tip at 5.4 K. Spectra over all three edges are qualitatively equivalent and featureless on the bare Au (see Supplementary Fig.~S4). Grey plots are the absolute value of the numerically calculated derivative of $\dd I/\dd V$; dashed lines in a) are Gaussian fits to the inelastic spin excitation signal. Steps at $\pm35\,\mathrm{mV}$ correspond to external vibrations of the CO molecule at the tip. \rev{Lock-in modulation: 3 and 2~mV~r.m.s, respectively}.
        c,d) Constant-height current Kondo maps of \1t and \2t, respectively (with a CO-functionalized tip, $V=2\,\mathrm{mV}$, scalebars $4\,\text{\si{\angstrom}}$, see also Supplementary Fig.~S7), compared with  simulated $\dd I/\dd V$ maps from the molecular Kondo orbitals (see Methods and \cite{CALVO-FERNANDEZ_Theoretical_2024})
        e,f) High-resolution magnetic field-dependent $\dd I/\dd V$ plots of the Kondo resonances at 320~mK of \1t and \2t, respectively. The red-dashed lines are Hurwitz-Fano fits of the zero-field curves \cite{JACOB_Temperature_2023,TURCO_Demonstrating_2024} providing intrinsic Kondo linewidths $\Gamma_K=(0.165\pm0.007)\,\mathrm{mV}$ and $\Gamma_K=(2.45\pm0.02)\,\mathrm{mV}$, which corresponds to $T_K=(0.49\pm0.02)\,\mathrm{K}$ and $T_K=(7.25\pm0.06)\,\mathrm{K}$ for \1t and \2t, respectively. \rev{Lock-in modulation: 0.2~mV~r.m.s.}
        }
 \end{figure*}

\noindent\textbf{Detection of $\pi$-magnetic ground state}\\
To unravel the magnetic properties of these nanographenes, we performed low-bias tunneling spectroscopy at the zigzag edges at the termini of both \1t and \2t molecules. The differential conductance ($\dd I/\dd V$) spectra in Figs.~\ref{fig:Fig3}a and \ref{fig:Fig3}b show \rev{narrow zero-bias peaks on both molecules, that we assign to} Kondo screening of unpaired $\pi$-electrons by the conduction electrons in the substrate~\cite{KONDO_Resistance_1964,LI_Single_2019}.
\rev{The line width of the Kondo resonances is similar over the three zig-zag edges of the molecules, indicating that variations in molecule-surface coupling are very small compared to the other relevant energy scales in this system  (Supplementary Fig.~S4).}

To resolve the origin of the Kondo state, we measured the evolution of the Kondo resonances under an applied magnetic field, at sub-Kelvin temperature. \rev{As shown in Fig.~\ref{fig:Fig3}e-f, at 0.32 K the Kondo peaks exhibit different intrinsic linewidth on each molecule, corresponding to Kondo temperature values of $T_K\!\sim\!0.5\,\mathrm{K}$ in the case of \1t and $T_K\!\sim\!7\,\mathrm{K}$ for \2t.
Increasing the magnetic field up to $8\,\mathrm{T}$ 
split the Kondo  peak only above fields of $2\,\mathrm{T}$ and $4\,\mathrm{T}$, respectively (Supplementary Fig.~S7b), persisting as a single resonance below these values. This lack of a peak split for low fields is characteristic of a (fully screened) spin-$\tfrac{1}{2}$ Kondo state ~\cite{LI_Single_2019,Sun_Maniplation_2022}, and rules out an underscreened Kondo effect (section S6), typically observed in nanographenes with higher spin~\cite{LI_Uncovering_2020,SU_Atomically_2020,WANG_AzaTriangulene_2022,VILAS-VARELA_OnSurface_2023,TURCO_Observation_2023}.}

As shown in the constant-height current maps in Fig.~\ref{fig:Fig3}c-d, the Kondo density of states is delocalized across the entire graphene platform, with enhanced intensity along the zigzag edges. This extended spatial distribution contrasts with the usual localization of $S = \tfrac{1}{2}$ Kondo resonances at a single radical center, suggesting the involvement of multiple spin-polarized orbitals. Simulations of Kondo amplitude maps performed within the multi-orbital Kondo theoretical framework of Ref.~\cite{CALVO-FERNANDEZ_Theoretical_2024}, also shown in Fig.~\ref{fig:Fig3}c-d, closely reproduce the experimental data. These results confirm that a multi-reference description is essential to capture the spin distribution in the ground state accurately (see Supplementary Figs.~S16 and S17 for the relevant Kondo orbitals for each species).

The low-energy spectrum of \1t\ exhibits additional features consisting of two bias-symmetric $\dd I/\dd V$ steps at $\sim\!\pm 100\,\mathrm{mV}$, which are absent in \2t\ (see Fig.~\ref{fig:Fig3}a and its numerical derivative). We attribute these steps to inelastic spin excitations induced by tunneling electrons~\cite{LI_Single_2019}. Given the expected multiradical character of this molecule and considering spin conservation, we assign them to a spin excitation from an $S = \tfrac{1}{2}$ ground state to an $S = \tfrac{3}{2}$ excited state. Similar to the Kondo resonance, the inelastic spin signal is delocalized across the entire molecular backbone in the maps shown in Supplementary Fig.~S9, with enhanced intensity at the zigzag edges. This spatial distribution is well-reproduced by the set of natural transition orbitals displayed in Supplementary Fig.~S8. In contrast, the larger nanographene \2t\ shows no resolvable spin inelastic features, even in spectra acquired at the lowest temperatures, suggesting excitation energies below our energy resolution.\\ 

 \begin{figure*}[t!]
        \begin{center}
   	\includegraphics[width=\textwidth]{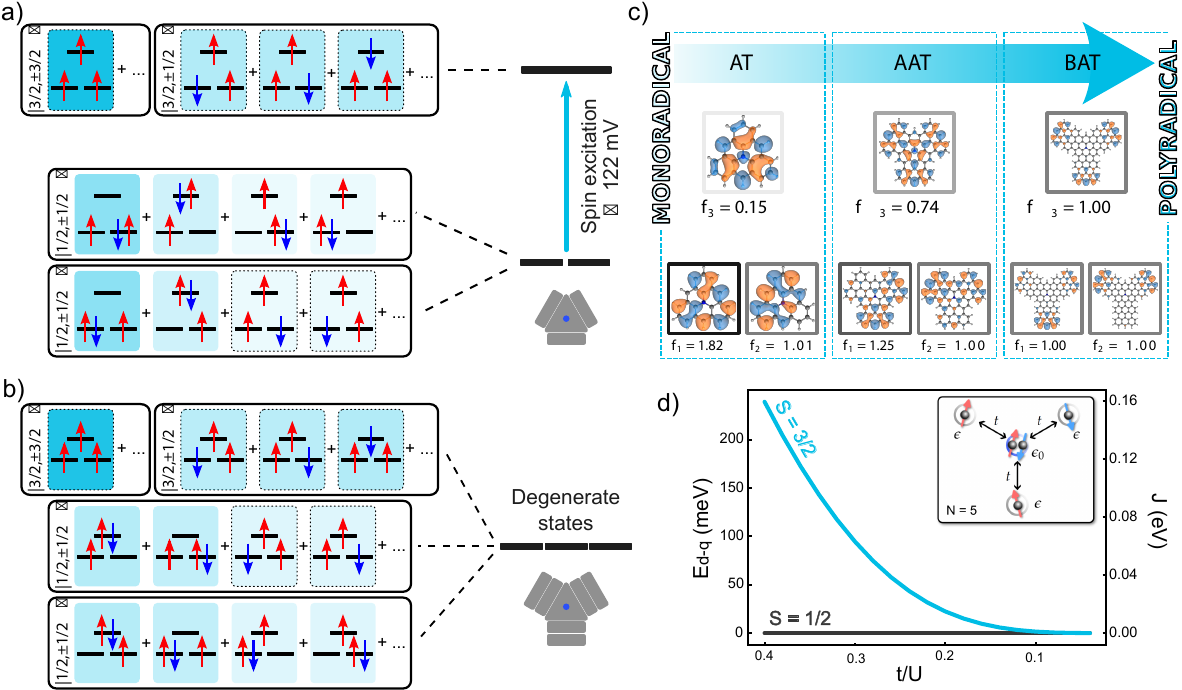}
        \caption{\textbf{Spin states and transition from mono to polyradical.} Many-body states diagram obtained from multi-reference CASSCF (11,11) calculations for a) \1t and b) \2t. It shows the most important single Slater determinant contributions to the three lowest energy states with the indicated total spin $|S,S_z\rangle$. The triradical Slater determinants appear highlighted by dashed lines. c) Scheme of the occupancy evolution of the three relevant natural orbitals as the size of the aza-nanographene increases, extracted from CASSCF(11,11). d) Energy difference between  doublet states and  excited quartet state \rev{and the corresponding magnetic exchange parameter ($\Delta E_{\text{d-q}}=3/2J$)~\cite{HARALDSEN_Neutron_2005}} as a function of the hopping parameter $t$ (in units of on site Coulomb repulsion $U$) obtained from an effective 4-site Hubbard model depicted as an inset (Supplementary Section~S15). }\label{fig:Fig4} 
        \end{center}
 \end{figure*}

\noindent\textbf{Frustrated spin trimer and spin triradical}\\
To describe the spin configurations of these nanographenes, we performed multi-reference quantum chemistry simulations within the Complete Active Space (CAS) framework \cite{SONG_Highly_2024,LI_Designer_2025,KUMAR_Multireference_2025}. 
We employed the complete active space self-consistent field (CASSCF) method with the N-electron valence state second-order perturbation theory (NEVPT2) correction included for capturing dynamic correlation effects outside the active space \cite{ANGELI_New_2007}. This approach enables a quantitative analysis of the spin states in the aza-nanographene platforms and their excitations. 

\rev{The multiconfigurational character of these systems becomes evident when analyzing the natural orbitals (NOs) obtained by diagonalizing the one-particle density matrix of the correlated CASSCF wavefunction  \cite{Davidson1972}. NOs provide a compact representation of correlated many-electron wavefunctions (Supplementary Section S11), and their fractional occupations directly quantify the degree of radical or polyradical character. }

\rev{The simulations shown in Fig.~\ref{fig:Fig4}c (complete representation in Supplementary Fig.~S15) show the progression of fractional occupation of the three active NOs found in the three smallest architectures ($n=0,1,2$). The covalent extension systematically reshapes the multiconfigurational electronic structure, transforming the monoradical \0t, into the triradical platforms \1t and  \2t.}
\rev{Rather than creating additional radical centers directly, anthene extensions reorganize the frontier natural orbitals NO1, NO2 and NO3, allowing active electrons to redistribute over spatially distinct edge-localized states. In \2t, the limiting case is reached and all three frontier natural orbitals become singly occupied and spatially separated. }

\rev{Despite the profound electronic reconfiguration with anthene extensions, the global  spin-$\tfrac{1}{2}$ ground state persists in all platforms, but with increasing ground state multiplicity. For \1t, the multi-reference simulations reproduce a two degenerate doublets and a low-energy spin-flip transition into a higher-spin quartet (Fig.~\ref{fig:Fig4}a), which is a hallmark of a frustrated spin trimer \cite{HARALDSEN_Neutron_2005}. The doublet–quartet excitation amounts to $\Delta E_\text{d-q}=122\,\mathrm{meV}$, in close agreement with the inelastic steps observed in the experimental spectra of \1t   (Fig.~\ref{fig:Fig3}a), validating the  frustrated polyradical ground state of \1t. 
The ground state of \2t emerges as a degenerate doublet–quartet manifold (Fig.~\ref{fig:Fig4}b), in agreement with the absence of inelastic spin excitations in the experiment (Fig.~\ref{fig:Fig3}b). This platform thus lies in a regime of effectively three non-interacting spins localized around  the zigzag edges of \2t, as mapped in Fig.~\ref{fig:Fig2}d, which give rise to spin-$\tfrac{1}{2}$ Kondo resonances in tunneling spectra (Supplementary Figs.~S4 and S8). }

\section*{\label{discussion}\raggedright DISCUSSION}\vspace{-0.5cm}
 
\noindent The electronic reconfiguration of the \0t core with anthene extension  is supported by the topological stabilization of edge radical states in the 7AGNR-like extensions, which progressively drives the edge-localization and decoupling of the frontier orbitals with increasing length~\cite{CAO_Topological_2017,WANG_Giant_2016,AJAYAKUMAR_pExtended_2022,LAWRENCE_Probing_2020}. 
Radical counting rules determine that a junction of three 7AGNR segments hosts a central junction state~\cite{TAMAKI_Topological_2020}, that here becomes doubly occupied by the extra electron of the aza moiety. 


Despite lying on the same sublattice, the three radicals in the \0t-$n$-anthene architectures exhibit an antiferromagnetic exchange interaction that vanishes with extension $n$. This spin coupling can be rationalized by modeling the spin trimers with a four-site Hubbard Hamiltonian populated by five electrons \rev{representing the topological edge and central states (Fig.~\ref{fig:Fig4}d). This  (5,4) model is the minimum one that accounts for the active space found in CAS simulations.  In its ground state, the model consistently yields an antiferromagnetic pattern rather than parallel Hund-like alignment \cite{JACOB_Theory_2022}, with a coupling strength that increases with the inter-site hopping $t$ (Supplementary Section S15).  
The origin of this antiferromagnetic exchange texture is the double occupation of the central site, which represents the aza core: modeling instead a pristine carbon system with a (4,4) model (Supplementary Fig.~S20b) reverses the correlations between terminal spins to a ferromagnetic-like pattern. } While Lieb’s theorem is strictly applicable only to half-filled lattices, it is notable that substitution of the central atom with nitrogen reverses the exchange pattern between terminal spins, stabilizing antiferromagnetic correlations.

These results establish  \2t\ as a prototypical molecular triradical  with topologically protected edge states and negligible exchange interactions, providing an ideal platform to explore correlated spin phenomena at the nanoscale. This spin trimer constitutes a molecular analog of the recently realized three-qubit system based on magnetic adatoms~\cite{WANG_atomicscale_2023}.  Benefiting from their weak exchange interaction, the spins of \2t\ can be selectively aligned by external or local magnetic fields and coherently addressed with microwaves, making \2t\ a promising platform for coherent spin manipulation.

\section*{\label{conclusion}\raggedright CONCLUSION}\vspace{-0.5cm}
\noindent  In summary, we present a molecular design strategy to reconfigure the number of radicals in an open-shell aza-triangulene nanographene by covalently incorporating anthene segments at its zigzag edges. This approach transforms a single-radical doublet into a frustrated spin triradical. The activation of additional radicals arises from the progressive stabilization of a topological phase within the attached anthene segments as their length increases, analogous to the behavior of short 7AGNRs, as revealed by multi-reference quantum calculations. Importantly, we demonstrate that nitrogen substitution in the aza-triangulene core reverses the exchange pattern, offering new chemical routes to control magnetic interactions at the molecular scale. By uniting chemical synthesis with the concepts of spin frustration and topological protection, our work establishes a general design principle for molecular platforms hosting symmetry-protected, weakly coupled spins. Such architectures provide a chemically programmable platform for quantum registers, paving the way toward the coherent manipulation of entangled spin states in graphene-based nanostructures.\\

\begin{acknowledgments}
The authors acknowledge financial support from grants PID2022-140845OBC61, PID2022-140845OBC62, PID2022-139776NB-C65, PID2023-146694NB-I00, and CEX2020-001038-M funded by MICIU/AEI/10.13039/501100011033 and the European Regional Development Fund (ERDF, A way of making Europe), from the FET-Open project SPRING (863098), the HE project HORIZON-EUROHPC-JU-2021- COE-01 01-101093374-MaX, the ERC Synergy Grant MolDAM (no. 951519), and the ERC-AdG CONSPIRA (101097693) funded by the European Union, 
from the European Union NextGenerationEU / PRTR-C17.I1 as well as by the IKUR Strategy of the Department of Education of the Basque Government, and from the Xunta de Galicia (Centro de Investigacion do Sistema Universitario de Galicia, 2023–2027, ED431G 2023/03). F.R.-L., and F.S. acknowledge funding by MICIU/AEI and FSE+ through scholarship No.  FPU20/03305, and through the Ramón y Cajal Fellowship RYC2021-034304-I, respectively. R.O. acknowledge financial suport within the scope of the project CICECO-Aveiro Institute of Materials, UIDB/50011/ 2020 (DOI 10.54499/UIDB/50011/2020), UIDP/50011/ 2020 (DOI 10.54499/UIDP/50011/2020), and LA/P/0006/ 2020 (DOI 10.54499/LA/P/0006/2020), financed by national funds through the FCT/MCTES (PIDDAC). P.J., M.K., and D.S. acknowledge the support of the Czech Science Foundation (GACR) project No.23-05486S and the CzechNanoLab Research Infrastructure supported by MEYSCR (LM2018110).
\end{acknowledgments}

\bibliography{refs_main}

\section*{\label{methods}\raggedright METHODS}\vspace{-0.5cm}
\noindent
\\
\noindent\textbf{Experimental setup}\\
The experiments were conducted in two different scanning tunneling microscope (STM) setups under ultrahigh vacuum conditions.
One is a home-built system that hosts a \textit{Createc GmbH} STM/AFM head with a base temperature of $5.4\,\mathrm{K}$. 
The second is a commercial $^3\mathrm{He}$ STM from \textit{Unisoku} operating at a base temperature of $321\,\mathrm{mK}$ and equipped with superconducting coils to induce either a magnetic field up to $9\,\mathrm{T}$ perpendicular to the sample surface, or a field up to $2\,\mathrm{T}$ in an arbitrary direction.
\\

\noindent\textbf{On-surface synthesis}\\
The Au(111) single crystal (\textit{MaTeck GmbH}) was prepared by successive cycles of Ne$^+$ or Ar$^+$ sputtering and annealing (550 $^\circ$C) for an atomically flat surface.
The molecular precursors were deposited by fast sublimation from a homemade flash silicon evaporator at high temperatures.
Alternatively, we deposited molecular precursor \textbf{1} in a controlled way from a Knudsen cell at 300 $^\circ$C using a commercial evaporator from \textit{Dodecon GmbH}.
We prepared samples including either precursor individually or both on the same surface.
After deposition, samples were typically annealed to around 330 $^\circ$C to induce the planarization reactions, as described in the main text. 
\\

\noindent\textbf{Imaging and spectroscopic measurements}\\
All STM and nc-AFM images were acquired under the specific conditions described in each figure caption and processed using the WSxM software~\cite{HORCAS_WSXM_2007}.
Bond-resolved imaging was typically performed using a CO-functionalized tip at close tip-sample distances, within the regime dominated by repulsive Pauli forces, operating in constant-height mode with bias voltages of less than 5 mV.
\rev{nc-AFM measurements were carried out with a qPlus~\cite{GIESSIBL_Highspeed_1998} quartz force sensor in the frequency modulation mode~\cite{Albrecht1991frequency} with a constant oscillation amplitude of 0.5\,\AA. The qPlus sensor had a resonance frequncy $f_0$ of $\sim$30.7\,kHz, quality factor $Q$ of $\sim$40k, and spring constant $k$ of $\sim$1.8\,kN/m.}

Scanning tunneling spectroscopy (STS) measurements were conducted in open-feedback mode under various conditions, employing a lock-in amplifier technique.

\rev{All key spectroscopic features were reproduced across multiple independently synthesized molecules and experimental runs. Molecular integrity was confirmed in each case by bond-resolved imaging (Supplementary Section S3), which limits statistical variation to inhomogeneous adsorption sites rather than structural defects. The broad IETS features of AAT were consistently reproduced across dozens of molecules. The Kondo linewidth of BAT exhibited some variations with the adsorption site, accompanied by minor asymmetries, attributed to local electrostatic gating near charging transitions.}

The presented spectra have been smoothed using a Savitzky–Golay filter, with window sizes ranging from 11 to 15 points and polynomial orders ranging from 3 to 5.
Kondo resonances were fitted using a Hurwitz–Fano function convoluted with the AC lock-in broadening, allowing extraction of the intrinsic Kondo width, as detailed in Ref.~\cite{JACOB_Temperature_2023,TURCO_Demonstrating_2024}. The temperature dependence of the intrinsic Kondo width can be fitted to extract the Kondo temperature. Ref.~\cite{TURCO_Demonstrating_2024} introduces an equation to extract the Kondo temperature from a single point, as we did in Fig.~\ref{fig:Fig3}, S2, and S3. \rev{Reported uncertainties in these values were derived from the standard error in the fit parameters.}
\\

\noindent\textbf{DFT and Multireference calculations}\\
The molecular geometries were optimized using density functional theory (DFT) in the high-spin state, as implemented in the FHI-AIMS software package~\cite{Blum2009}, employing the PBE0 hybrid functional~\cite{Adamo1999}. In these calculations, the Tkatchenko-Scheffler method was used to account for Van der Waals interactions. Given the open-shell and multi-radical nature of the molecules, the complete active space configuration interaction (CASCI) method was employed to obtain an accurate description of the wave function and electronic energies. One- and two-electron integrals are constructed in the basis of molecular orbitals around the Fermi energy using the PBE functional, which were derived using the quantum chemistry software ORCA~\cite{Neese2012} using the orbitals from the restricted open-shell Kohn-Sham (ROKS). Natural orbitals (NOs) are obtained by diagonalizing the one-particle reduced density matrix constructed from the ground state of the ground-state multireference CASCI wavefunction.
\\
\textbf{Natural Transition Orbitals (NTO):} To simulate the d$I$/d$V$ maps corresponding to IETS spin excitation maps, we have calculated the Natural Transition Orbitals (NTOs)~\cite{Martin2003}, which correspond to the electronic transition density matrix of the single spin flip process from the first and second doublet ground state to the quartet excited states.
\\
\textbf{Dyson Orbitals:} To accurately interpret experimental differential conductance (d$I$/d$V$) maps for these molecules exhibiting significant multireference character, we have calculated the Dyson orbitals~\cite{Ortiz2020a,KUMAR_Multireference_2025, Zuzak2024}. They represent the overlap between the many-body wavefunctions before and after electron addition or removal, capturing the spatial distribution of the states directly involved in tunneling. 
\\
\textbf{Kondo Orbitals (KO):} Kondo orbitals are calculated by diagonalizing the Hamiltonian derived from the multi-channel Anderson model, which considers the many-body multiplet structure of molecules obtained from the CASCI calculation for the neutral ground state and virtual charge states as described in the Ref.~\cite{CALVO-FERNANDEZ_Theoretical_2024}.
\\
\textbf{Theoretical d$I$/d$V$ maps of NTOs, Kondo, and Dyson orbitals}  were calculated using the Probe Particle Scanning Probe Microscopy (PP-SPM) code \cite{Krejci2017}.

\setcounter{figure}{0}
\renewcommand{\figurename}{\textbf{Extended Data Fig.}}
\renewcommand{\thefigure}{\textbf{\arabic{figure}}}

\end{document}


\title{Supplementary Information: \\
Topological Engineering of a Frustrated Antiferromagnetic Triradical in Aza-Triangulene Architectures}

\author{Francisco Romero-Lara}\thanks{These authors contributed equally to this work.}
    \affiliation{CIC nanoGUNE-BRTA, 20018 Donostia-San Sebasti\'an, Spain}
\author{Manuel Vilas-Varela}\thanks{These authors contributed equally to this work.}
    \affiliation{Centro Singular de Investigaci\'on en Qu\'imica Biol\'oxica e Materiais Moleculares (CiQUS) and Departamento de Qu\'imica Org\'anica, Universidad de Santiago de Compostela, 15782 Santiago de Compostela, Spain}
\author{Manish Kumar}\thanks{These authors contributed equally to this work.}
    \affiliation{Institute of Physics of the Czech Academy of Sciences, Cukrovarnicka 10, Prague 6, CZ 16200, Czech Republic}
\author{Ricardo Ortiz}\thanks{These authors contributed equally to this work.}
    \affiliation{Donostia International Physics Center (DIPC), 20018 Donostia-San Sebastian, Spain}
    \affiliation{CICECO- Aveiro Institute of Materials, Department of Chemistry, University of Aveiro, 3810-193 Aveiro, Portugal}
\author{Alessio Vegliante}
    \affiliation{CIC nanoGUNE-BRTA, 20018 Donostia-San Sebasti\'an, Spain}
\author{Lucía Gómez-Rodrigo}
    \affiliation{Centro Singular de Investigaci\'on en Qu\'imica Biol\'oxica e Materiais Moleculares (CiQUS) and Departamento de Qu\'imica Org\'anica, Universidad de Santiago de Compostela, 15782 Santiago de Compostela, Spain}
\author{Trisha Sai}
    \affiliation{CIC nanoGUNE-BRTA, 20018 Donostia-San Sebasti\'an, Spain}
\author{Jan Patrick Calupitan}
    \affiliation{Donostia International Physics Center (DIPC), 20018 Donostia-San Sebastian, Spain}
    \affiliation{Centro de F\'isica de Materiales (CFM-MPC), Centro Mixto CSIC-UPV/EHU, E-20018 Donostia-San Sebasti\'an,  Spain}
    \affiliation{Sorbonne Université, CNRS, Institut Parisien de Chimie Moléculaire (IPCM), F-75005 Paris, France}
\author{Diego Soler}
    \affiliation{Institute of Physics of the Czech Academy of Sciences, Cukrovarnicka 10, Prague 6, CZ 16200, Czech Republic}
\author{Niklas Friedrich}
    \affiliation{CIC nanoGUNE-BRTA, 20018 Donostia-San Sebasti\'an, Spain}
\author{Dongfei Wang}
    \affiliation{CIC nanoGUNE-BRTA, 20018 Donostia-San Sebasti\'an, Spain}
    \affiliation{Key Laboratory of Advanced Optoelectronic Quantum Architecture and Measurement, School of Physics, Beijing Institute of Technology, 100081 Beijing, China}
\author{Jon Ortuzar}
    \affiliation{CIC nanoGUNE-BRTA, 20018 Donostia-San Sebasti\'an, Spain}
\author{Leonard Edens}
    \affiliation{CIC nanoGUNE-BRTA, 20018 Donostia-San Sebasti\'an, Spain}
\author{Stefano Trivini}
    \affiliation{CIC nanoGUNE-BRTA, 20018 Donostia-San Sebasti\'an, Spain}
\author{Thomas Frederiksen}
    \affiliation{Donostia International Physics Center (DIPC), 20018 Donostia-San Sebastian, Spain}
    \affiliation{Ikerbasque, Basque Foundation for Science, 48009 Bilbao, Spain}
\author{Fabian Schulz}
    \affiliation{CIC nanoGUNE-BRTA, 20018 Donostia-San Sebasti\'an, Spain}
\author{Pavel Jelínek}\email{jelinekp@fzu.cz}
    \affiliation{Institute of Physics of the Czech Academy of Sciences, Cukrovarnicka 10, Prague 6, CZ 16200, Czech Republic}
\author{Diego Peña}\email{diego.pena@usc.es}
    \affiliation{Centro Singular de Investigaci\'on en Qu\'imica Biol\'oxica e Materiais Moleculares (CiQUS) and Departamento de Qu\'imica Org\'anica, Universidad de Santiago de Compostela, 15782 Santiago de Compostela, Spain}
    \affiliation{Oportunius, Galician Innovation Agency (GAIN), 15702 Santiago de Compostela, Spain}
\author{José Ignacio Pascual}\email{ji.pascual@nanogune.eu}
    \affiliation{CIC nanoGUNE-BRTA, 20018 Donostia-San Sebasti\'an, Spain}
    \affiliation{Ikerbasque, Basque Foundation for Science, 48009 Bilbao, Spain}

\date{\today}
\maketitle

\tableofcontents

\makeatletter 
\renewcommand{\thefigure}{S\@arabic\c@figure}
\renewcommand{\thesubsection}{S\arabic{subsection}}
\renewcommand\subsection{\@startsection
  {subsection}{2}{-0.5em}
  {-2.0ex \@plus -1ex \@minus -.2ex}
  {1.0ex \@plus .2ex}
  {\Large\bfseries} 
}
\def\@makecaption#1#2{%
  \par\vskip\abovecaptionskip
  \begingroup
  \fontsize{11pt}{13pt}\selectfont 
  \leftskip=0pt\rightskip=0pt
  \noindent #1.\ #2\par
  \endgroup
  \vskip\belowcaptionskip}
\makeatother

\onecolumngrid

\newpage

\subsection{In solution synthesis of molecular precursors}

\noindent Starting materials were purchased reagent grade from TCI and Sigma-Aldrich and used without further purification. Trimethyl 2,2',2''-nitrilotribenzaldehyde (\textbf{3}) was synthesized following a reported procedure~\cite{VILAS-VARELA_OnSurface_2023}. Organolithium reagents (\textbf{4} and \textbf{5}) were prepared by treatment of the corresponding halides with n-butyllithium~\cite{MATEO_OnSurface_2020}. All reactions were carried out in oven-dried glassware, which was cooled down under an inert atmosphere of purified Ar using Schlenk techniques. Thin-layer chromatography (TLC) was performed on Silica Gel 60 F-254 plates (Merck). Column chromatography was performed on silica gel (40-60 µm). NMR spectra were recorded on a Bruker Varian Inova 500 spectrometer.
\\

\noindent\textbf{Synthesis of precursor \textbf{1} (\1t)} \\

\noindent Over a solution of the lithium derivative \textbf{4} (2.30 mmol, 5.00 equiv.) in THF (20 mL), aldehyde \textbf{3} (150 mg, 0.46 mmol) was added at 0$^\circ$C.
After the addition, the mixture was allowed to reach room temperature and was stirred for 16 h.
Then, \ce{MeOH} (2 mL) was added, and the solvent was evaporated under reduced pressure.
The residue was redissolved in \ce{CH2Cl2} (25 mL), and \ce{BF3}·\ce{OEt2} (1.00 mL) was added at 0$^\circ$C.
After the addition, the mixture was stirred for 30 min and poured into an aqueous \ce{NaOH} solution (2 M, 30 mL).
After separating phases, the organic extract was dried over anhydrous \ce{Na2SO4}, filtered, and evaporated under reduced pressure.
The residue was purified by column chromatography (\ce{SiO2}; hexane:\ce{CH2Cl2} 3:2), affording compound \textbf{1} (38 mg, 10\%) as a yellow solid.

\textbf{$^1$H NMR} (500 MHz, \ce{CDCl3}) $\updelta$: 8.67 – 8.55 (m, 2H), 8.49 (m, 1H), 8.10 (m, 1H), 8.03 – 7.97 (m, 1H), 7.62 (m, 1H), 7.56 – 7.50 (m, 1H), 7.46 – 7.35 (m, 3H), 6.46 – 6.39 (m, 2H), 6.34 – 6.27 (m, 1H) ppm.

\textbf{$^{13}$C NMR-DEPT} (125 MHz, \ce{CDCl3}) $\updelta$: 139.08 (C), 138.79 (C), 133.11 (C), 133.07 (C), 132.85 (C), 132.66 (C), 132.08 (C), 132.05 (C), 131.32 (C), 129.69 (CH), 129.66 (CH), 129.54 (CH), 129.48 (CH), 129.39 (C), 129.30 (C), 128.27 (CH), 128.16 (CH), 128.13 (CH), 128.02 (CH), 127.99 (CH), 127.45 (CH), 127.38 (CH), 126.85 (CH), 126.79 (CH), 125.79 (CH), 125.66 (C), 125.34 (C), 125.19 (C), 125.15 (CH), 125.11 (CH), 124.89 (CH), 124.85 (CH), 123.58 (CH), 123.44 (CH), 123.23 (CH), 123.05 (CH), 39.03 (CH) ppm.

\textbf{MS (MALDI-TOF)} for \ce{C63H39N}; calculated: 809.31, found: 809.28.
\\

\noindent\textbf{Synthesis of precursor \textbf{2} (\2t)}\\

\noindent Over a solution of the lithium derivative \textbf{5} (0.55 mmol, 4.00 equiv) in THF (15 mL), aldehyde \textbf{3} (46 mg, 0.14 mmol) was added at 0$^\circ$C.
After the addition, the mixture was allowed to reach room temperature and was stirred for 16 h.
Then, water (10 mL) was added, and the solvent was evaporated under reduced pressure.
The residue was redissolved in \ce{CH2Cl2} (20 mL), and \ce{BF3}·\ce{OEt2} (700 $\upmu$L) was added at 0$^\circ$C.
After the addition, the mixture was stirred for 30 min and poured into an aqueous NaOH solution (2 M, 50 mL).
After separating phases, the organic extract was dried over anhydrous \ce{Na2SO4}, filtered, and evaporated under reduced pressure.
The residue was purified by column chromatography (\ce{SiO2}; hexane:\ce{CH2Cl2} 3:2), affording compound \textbf{2} (39 mg, 21\%) as a yellow solid.

\textbf{$^1$H NMR} (500 MHz, \ce{CDCl3}) $\updelta$: 8.93 – 8.79 (m, 2H), 8.70 (m, 1H), 8.22 – 8.10 (m, 2H), 7.71 (m, 1H), 7.62 (m, 1H), 7.50 – 7.39 (m, 3H), 7.25 – 7.18 (m, 4H), 7.18 – 7.07 (m, 3H), 7.03 (m, 1H), 6.84 (m, 1H), 6.76 (m, 1H), 6.68 (m, 1H) ppm.

\textbf{$^{13}$C NMR-DEPT} (125 MHz, \ce{CDCl3}) $\updelta$: 139.65 (C), 134.34 (C), 134.29 (C), 133.69 (C), 133.58 (C), 133.16 (C), 133.12 (C), 132.99 (C), 132.88 (C), 131.96 (C), 131.86 (C), 131.82 (C), 131.81 (C), 131.74 (C), 131.73 (C), 131.69 (C), 131.22 (C), 131.18 (C), 129.23 (C), 129.08 (C), 128.73 (CH), 128.67 (CH), 128.54 (C), 128.43 (CH), 128.30 (CH), 128.26 (CH), 128.18 (CH), 127.42 (CH), 127.36 (CH), 127.16 (CH), 127.06 (CH), 126.99 (CH), 126.18 (C), 126.12 (CH), 126.08 (CH), 125.95 (CH), 125.80 (CH), 125.76 (C), 125.60 (CH), 125.56 (CH), 125.53 (CH), 125.45 (CH), 125.40 (CH), 125.33 (CH), 123.88 (CH), 123.71 (CH), 123.62 (CH), 123.58 (CH), 39.36 (CH) ppm.

\textbf{MS (MALDI-TOF)} for \ce{C105H63N}; calculated: 1337.50, found: 1337.43.

 \begin{figure}[!htb]
        \centering
    	\includegraphics[width=0.9\textwidth]{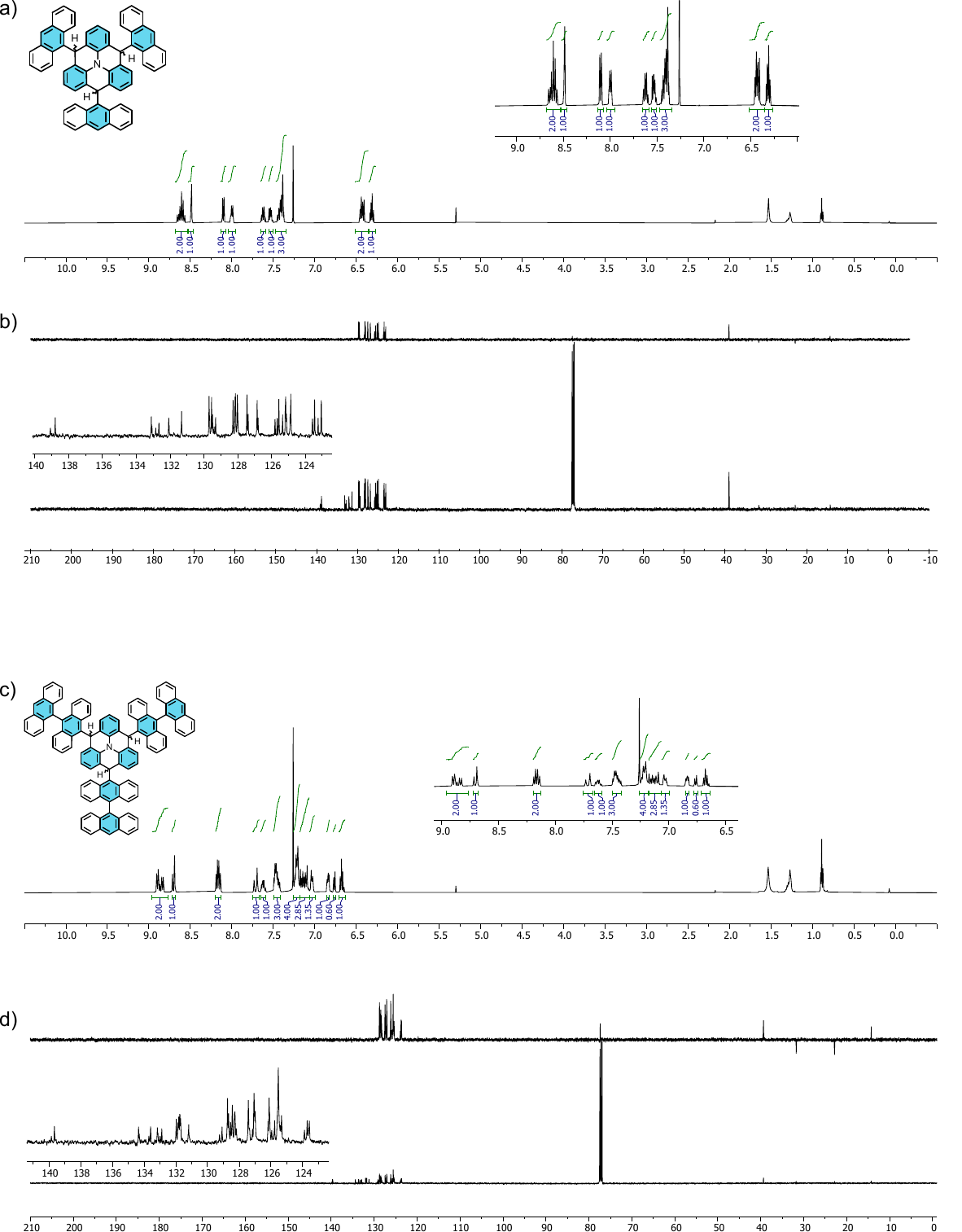}
        \caption{$^1$H NMR (500 MHz, \ce{CDCl3}) spectrum of compound a) \textbf{1} and c) \textbf{2}. $^{13}$C NMR-DEPT (125 MHz, \ce{CDCl3}) spectrum of compound a) \textbf{1} and d) \textbf{2}.}
        \label{fig:FigS_chem}
 \end{figure}

\newpage





\subsection{Frequently observed fragments}

\begin{figure}[h!]
        \centering
    	\includegraphics[width=\textwidth]{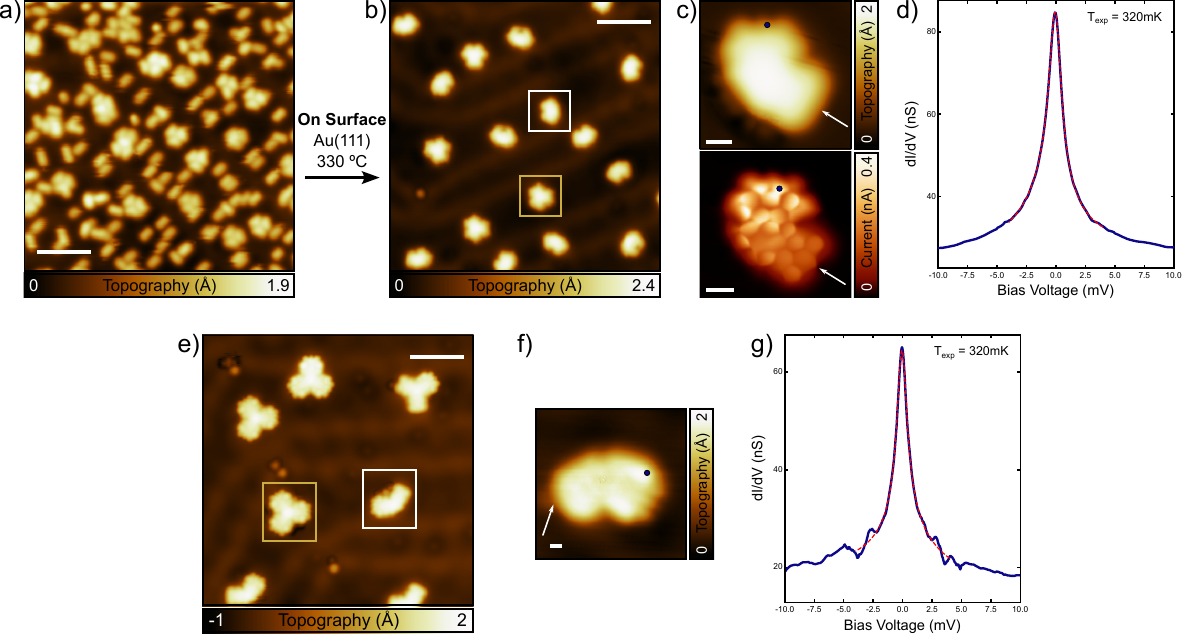}
        \caption{a) Image of the sample showing several unreacted precursors \textbf{1} and a large amount of anthracenes, products of the fragmentation upon sublimation. b) and e) Overview image of the sample after the reaction on-surface for \1t and \2t, respectively. Yellow squares highlight the targeted nanographenes, while white squares highlight the twofold fragments. c) STM (top) and BR-STM (bottom) images of a twofold fragment of \1t. White arrows point to extra hydrogen. d) and g) High-resolution STS spectrum measured over the blue dots in c) and f) for \1t and \2t hydrogenated fragments respectively. The red-dashed lines in the spectra are Hurwitz-Fano fits yielding a Kondo temperature~\cite{JACOB_Temperature_2023,TURCO_Demonstrating_2024} of $T_K=(3.09\pm0.01)\,\mathrm{K}$ and $T_K=(2.29\pm0.04)\,\mathrm{K}$ for \1t and \2t hydrogenated fragments respectively. f) STM image of a twofold fragment of \2t. A white arrow points to an extra hydrogen. a) and b): scalebars 4 nm, I$=50$ pA, V$=200$ mV; c): scalebars 4 \si{\angstrom}, I$=50$ pA, V$=200$ mV (top), V$=5$ mV (bottom); e): scale 4 nm, I$=30$ pA, V$=100$ mV; f): scalebar 4 nm, I$=30$ pA, V$=5$ mV.}
        \label{fig:FigS_frag}
 \end{figure}

\noindent Although our primary focus remains on the intact nanographenes, we frequently encountered twofold symmetric structures (highlighted by the white squares in Fig.~\ref{fig:FigS_frag}b and Fig.~\ref{fig:FigS_frag}e).
We can attribute them to the product of a fragmented precursor.
Precursors \textbf{1} and \textbf{2} occasionally lose one or more anthracene extensions upon deposition on the surface.
A proof of this fragmentation can be seen in Fig.~\ref{fig:FigS_frag}a.
Threefold structures corresponding to the unreacted precursor and smaller elongated structures corresponding to the fragmented anthracenes can be seen.
Interestingly, the fragments can host unpaired electrons and readily capture additional hydrogen atoms as seen in Fig.~\ref{fig:FigS_frag}c,f.
Such hydrogenated fragments serve as single-radical spin-1/2 platforms, hence the intense Kondo resonance measured over the unquenched edge (Fig.~\ref{fig:FigS_frag}d,g).

\newpage

\subsection{Tip-induced removal of extra H-atoms}

\begin{figure}[th!]
        \centering
    	\includegraphics[width=\textwidth]{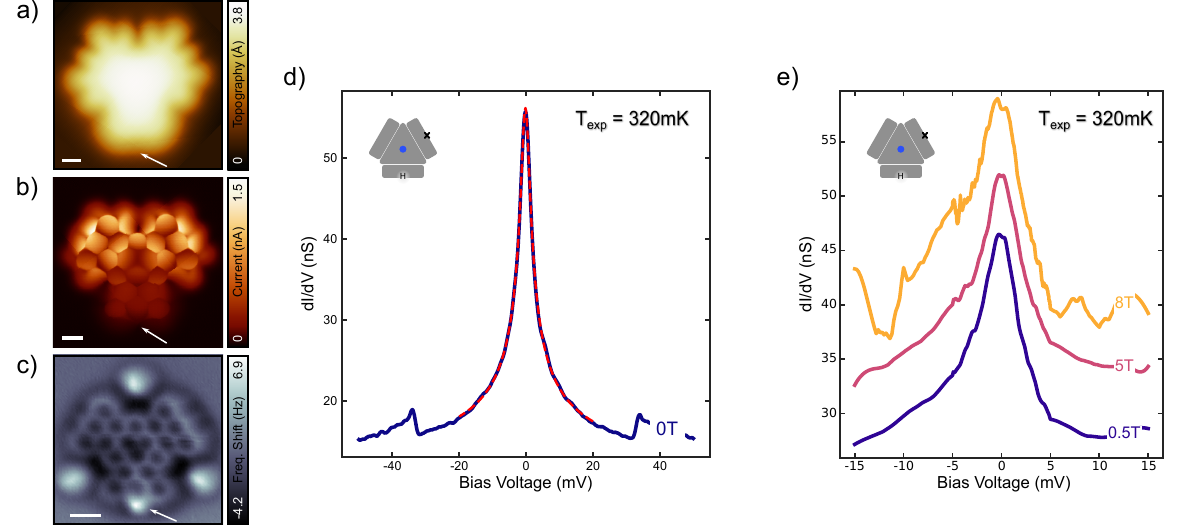}
        \caption{a) STM b) BR-STM and c) nc-AFM images of \1t, respectively, with the bottom zigzag edge quenched by an extra hydrogen atom (white arrow). \rev{d) High-resolution STS spectrum measured over one of the unquenched edges of the \textbf{1H-}\1t at $0\,\mathrm{T}$ using a CO tip.
        The red-dashed line in the spectrum is a Hurwitz-Fano fit yielding a Kondo temperature~\cite{JACOB_Temperature_2023,TURCO_Demonstrating_2024} of $T_K=(10.47\pm0.07)\,\mathrm{K}$. g) Magnetic field-dependent high-resolution STS spectra of \textbf{1H-}\1t using a metallic tip.}
        Setpoints of the images are a): I$=100$ pA, V$=10$ mV; b): V$=5$ mV; c): V$=0$ mV; Scalebars: 4 \si{\angstrom}.}
        \label{fig:FigS_extraH}
 \end{figure}

We occasionally encountered molecules with additional hydrogen atoms bonded to carbons at the zigzag edges, effectively quenching radical electrons (Fig.~\ref{fig:FigS_extraH}a-c).
Both nc-AFM and STM can detect this.
Specifically, the extra hydrogen atom yields a brighter contrast in nc-AFM due to the change in hybridization of the carbon from $sp^2$ to $sp^3$, as in Fig.~\ref{fig:FigS_extraH}c.
Conversely, the BR-STM image of the zigzag edge where the radical has been quenched shows a darker contrast, as in Fig.~\ref{fig:FigS_extraH}b.
This contrast arises due to the decreased density of states near the Fermi level on this particular edge.
These hydrogen atoms can be easily removed through high-energy electron tunneling. 

\rev{The hydrogenated \1t in Fig.~\ref{fig:FigS_extraH}a-c undergoes charge transfer, similarly to \0t. This results in a spin-1/2 radical rather than the expected singlet state of the neutral species, as predicted by CASSCF calculations. In that case, a singlet-triplet spin excitation is expected at $\sim\!80\,\mathrm{meV}$, which is not observed experimentally.
In contrast, the hydrogenated molecule shows an intense Kondo resonance (Fig.~\ref{fig:FigS_extraH}d), characteristic of spin-1/2 nanographenes.
The Kondo peak starts splitting only under an $8\,\mathrm{T}$ magnetic field (Fig.~\ref{fig:FigS_extraH}e), as opposed to the underscreened case, where a Zeeman-splitting should be detected above $0.5\,\mathrm{T}$ for this experimental temperature.}

\clearpage

\subsection{Comparison of STS spectra over each zigzag edge} 

 \begin{figure}[th!]
        \centering
    	\includegraphics[width=\textwidth]{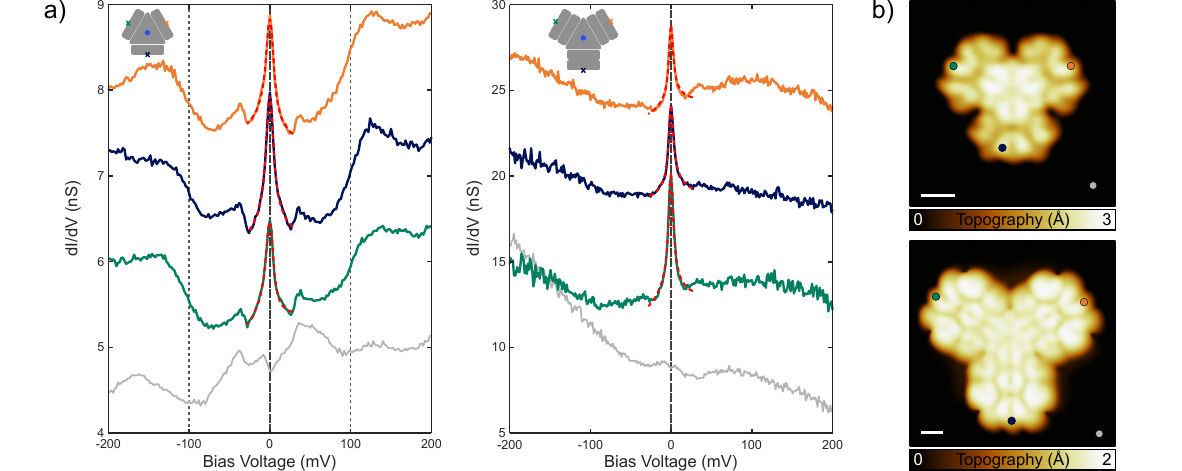}
        \caption{a) STS spectra measured at $5.4\,\mathrm{K}$ over each zigzag edge and on the bare substrate for \1t (left) and \2t (right) as indicated by the colored dots in b). Spectra were shifted vertically for clarity, and dashed lines were added as visual guides for Kondo and spin excitation.
        The inelastic fingerprints of the spin excitation are only present in \1t.
        Red-dashed lines are Hurwitz-Fano fits yielding intrinsic Kondo widths~\cite{JACOB_Temperature_2023,TURCO_Demonstrating_2024} of $\Gamma_K=5.97\,\mathrm{mV}$, $\Gamma_K=5.52\,\mathrm{mV}$, $\Gamma_K=5.52\,\mathrm{mV}$ for orange, blue and green respectively in \1t; and $\Gamma_K=3.62\,\mathrm{mV}$, $\Gamma_K=4.20\,\mathrm{mV}$, $\Gamma_K=4.33\,\mathrm{mV}$ for orange, blue and green respectively in \2t.
        Spectra on different edges show insignificant variations of less than $1\,\mathrm{mV}$ in their Kondo linewidth.
        b) High-resolution CO-STM images of \1t (top, $I=50\,\mathrm{pA}$, $V=5\,\mathrm{mV}$) and \2t (bottom, $I=20\,\mathrm{pA}$, $V=2\,\mathrm{mV}$). Scalebars 4 \si{\angstrom}.}
        \label{fig:FigS_Kondo}
 \end{figure}

\clearpage

\subsection{Kelvin probe force microscopy measurements demonstrating the neutral charge state of AAT and BAT}

\begin{figure}[b!]
    \centering
    \includegraphics[width=\textwidth]{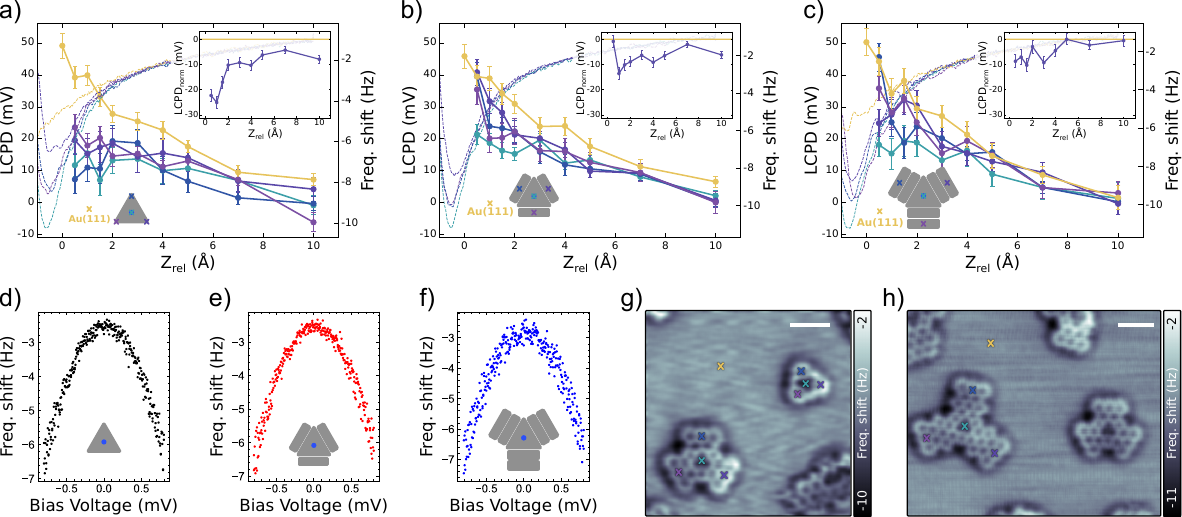}
    \caption{\textbf{LCPD measurements for AT, AAT, and BAT on Au(111).} a-c) Local contact potential difference (LCPD) as a function of relative tip-sample distance measured over the three molecules as indicated in panels g) and h). Insets: $\mathrm{LCPD}\,(z)$ averaged over the different molecular sites plotted relative to the values on Au(111). d-f) Exemplary $\Delta \text{f}(V)$ curves for the three molecules, showing the expected parabolic behaviour. g-h) Bond-resolved nc-AFM images of \0t, \1t, and \2t, indicating the positions where the $\Delta \text{f}(V)$ spectra were measured (colored crosses). All data were acquired with the same CO-functionalized tip, at an oscillation amplitude of 0.5\,\AA. For LCPD measurements, the bias voltage was swept between $\pm$800\,mV and maximum currents did not exceed 1\,nA. Z$_\text{rel}$=0 refers to opening of the feedback loop over the center of the molecule at a setpoint of $V$=5\,mV, $I$=10\,pA. Positive Z$_\text{rel}$ values refer to an increase of the tip-sample distance. 
    }
    \label{fig:FigS_KPFS}
\end{figure}

Here, we provide evidence for the charge-neutral ground state of \1t and \2t on Au(111) through measurements of the local contact potential difference (LCPD) using nc-AFM~\cite{Gross2009_chargestateAFM}. In Fig.~\ref{fig:FigS_KPFS}a-c, we compare the LCPD of \1t and \2t with that of the smaller aza-[3]triangulene \0t, which is known to lie in a cationic ground state on Au(111) due to electron transfer to the substrate~\cite{WANG_AzaTriangulene_2022}. To that end, with the tip either over the molecule or the Au(111) substrate, we recorded the frequency shift $\Delta \text{f}$ as a function of bias voltage for different relative tip-sample distances Z$_\text{rel}$ over a range of 1 nm. From a parabolic fit to the $\Delta \text{f}(V)$ curves, we extracted their LCPD (voltage at the maximum of the parabola) and plotted these values as a function of tip-sample distance, $\mathrm{LCPD}\,(z)$. Importantly, all data were acquired with the exact same CO-functionalized tip, allowing for qualitative and quantitative comparison of the observed trends. The exemplary $\Delta \text{f}(V)$ curves in Fig.~\ref{fig:FigS_KPFS}d-f confirm that the frequency shift exhibits the expected parabolic dependence on the bias voltage, with no signatures of tip-induced charging of either of the three molecules.

As shown in Fig.~\ref{fig:FigS_KPFS}a, the $\mathrm{LCPD}\,(z)$ on \0t shifts towards more negative values from the reference LCPD on the bare substrate when the tip approaches closer than $Z_\text{rel}=3\,\mathrm{\AA}$. This trend reflects its well-known cationic charge state on Au(111), and becomes more pronounced as the tip further approaches the \0t molecule. In contrast, for \1t and \2t the evolution of $\mathrm{LCPD}\,(z)$ (Figs.~\ref{fig:FigS_KPFS}b and \ref{fig:FigS_KPFS}c, respectively) is comparable to that on the Au(111) substrate for equivalent distances. These different trends, for \0t on the one hand and \1t, \2t on the other hand, are even more evident in the insets in Fig.~\ref{fig:FigS_KPFS}a-c, which compare the deviation of $\mathrm{LCPD}\,(z)$ from the Au(111) reference, averaged over several points on each molecule. The pronounced shift toward more negative LCPD values on \0t contrasts with the nearly constant deviation for \1t and \2t, in agreement with the assignment of \1t and \2t being charge-neutral on Au(111).

\begin{figure}[th!]
    \centering
    \includegraphics[width=\textwidth]{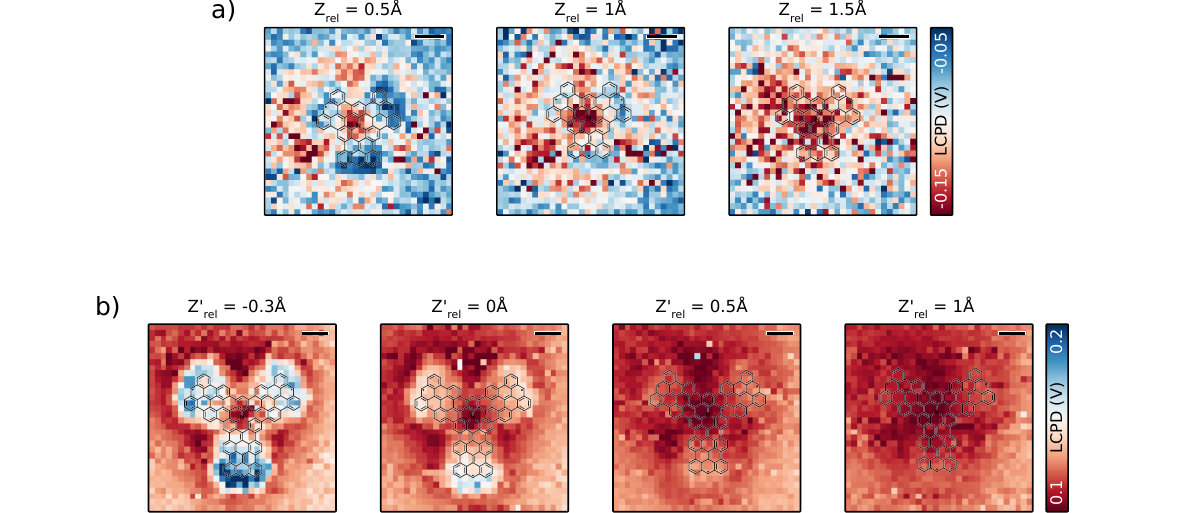}
    \caption{\textbf{LCPD maps at different tip–sample distances.} a, b) Spatially resolved maps of the LCPD recorded for \1t (a) and \2t (b). The same color scales are used for all heights. Maps for \1t and \2t were acquired with two different CO-functionalized tips. Relative tip–sample distance $Z_\text{rel}$=0 corresponds to the setpoint of $10\,\mathrm{mV}$, $10\,\mathrm{pA}$ over the nitrogen position for \1t and $Z\text{'}_\text{rel}$=0 to the setpoint of $600\,\mathrm{mV}$, $1\,\mathrm{nA}$ over the nitrogen position for \2t. Positive (negative) Z$_\text{rel}$ values refer to an increase (decrease) of the tip-sample distance with respect to that setpoint. Scalebars: 5\,\AA.
    }
    \label{fig:FigS_LCPD}
\end{figure}

Figures~\ref{fig:FigS_KPFS}b and \ref{fig:FigS_KPFS}c also show that the LCPD value reduces significantly at small $Z_\text{rel}$ tip positions over the aza center on \1t and \2t. This reveals a local positive partial charge at the N site arising from $\pi$-electron redistribution over the whole molecule. This is further supported by spatial mapping of the LCPD signal over the \1t and \2t molecules at constant $Z_\text{rel}$ (Fig.~\ref{fig:FigS_LCPD}). At larger tip–sample distances, the LCPD over the molecules is similar to, or slightly lower than, that of the surface, which we attribute to the push-back (pillow) effect at the metal–molecule interface~\cite{Crispin_Characterization_2002}. At closer distances, still within the attractive regime, the LCPD becomes locally more positive on both molecules, and its spatial distribution reproduces an electronic redistribution from the aza center towards the radical sites. This redistribution correlates with the LDOS close to the Fermi energy and explains why the LCPD becomes locally more positive at the radical sites, even though the adsorbate remains globally neutral. The KPFM maps were acquired with different CO-functionalized STM tips, so the absolute LCPD values cannot be compared quantitatively; however, the distance dependence and the trends relative to the bare substrate provide a robust basis for qualitative comparison.

\clearpage

\subsection{Magnetic field-dependent STS analysis}

 \begin{figure}[b!]
        \centering
    	\includegraphics[width=\textwidth]{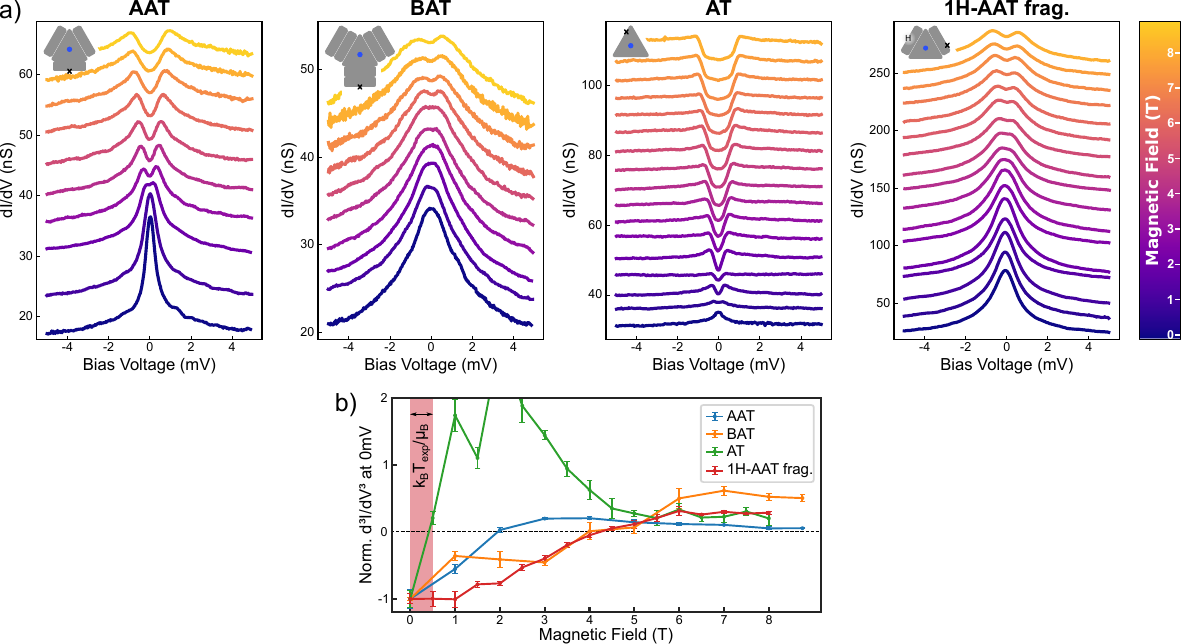}
        \caption{a) High-resolution STS spectra of \1t, \2t, \0t, and \textbf{1H-AAT} twofold fragment (Fig.~\ref{fig:FigS_frag}c,d), as a function of magnetic field. Spectra are vertically shifted for clarity. b) Computed curvature of the Kondo peaks at $V=0$, $\dd^3I/\dd V^3(B,V=0)$, normalized by the value at zero field, $\dd^3I/\dd V^3(B=0,V=0)$. 
        }
        \label{fig:FigS_KondoBfield}
 \end{figure}

\noindent The splitting of the Kondo resonance under an applied magnetic field is a well-established phenomenon. Early numerical renormalization group (NRG) calculations by Costi  \cite{COSTI_Kondo_2000}, later refined by other methods  \cite{Rosch_Spectral_2003,Filippone_At_2018}, found that a $\mathrm{S}\!=\!1/2$ impurity is expected to develop a splitting of the zero-bias Kondo state for magnetic fields in the order of the Kondo temperature, e.g. $\mathrm{B}(\mathrm{T}\!=\!0)\!\approx\! k_BT_K/\mu_B$, being $k_B$ and $\mu_B$ the Boltzmann constant and the Bohr magneton, respectively. For weak fields, the response is only a broadening and lowering of its zero-bias conductance peak. To identify the onset of splitting, a sign change in the curvature of the Kondo resonance at $V\!=\!0$ is commonly plotted.

In contrast, nanographenes with larger spin on a surface are usually partially screened or underscreened on the surface, exhibiting weaker Kondo states that behave differently under magnetic fields from the fully screened (Fermi liquid) scenario.   As shown previously for $\mathrm{S}\!>\!1/2$ nanographenes \cite{LI_Uncovering_2020,VILAS-VARELA_OnSurface_2023,SU_Atomically_2020}, the underscreened Kondo resonance splits into Zeeman states at a much lower field than in the fully screened case; in particular, the observation of the splitting is limited by the temperature of the measurement. In essence, the appearance of a Zeeman-split Kondo peak at a given field value will determine the screening regime and, more importantly, the $\mathrm{S}\!=\!1/2$ or higher spin ground state.



In Fig.~\ref{fig:FigS_KondoBfield}a, we compare the magnetic field response of \1t and \2t with that of \0t molecules and a hydrogenated \1t fragment (Fig.~\ref{fig:FigS_frag}c), as a reference, all measured at $320\,\mathrm{mK}$ and with magnetic fields up to $8.75\,\mathrm{T}$. As shown earlier~\cite{WANG_AzaTriangulene_2022}, \0t lies in a cationic $\mathrm{S}\!=\!1$ ground state on Au(111) due to charge transfer, and exhibits a weak underscreened Kondo resonance.  
In Fig.~\ref{fig:FigS_KondoBfield}b, we compare the computed curvature of the Kondo peaks at $V\!=\!0$ (i.e., $\dd^3I/\dd V^3(B,V\!=\!0)$) for the four molecules, to determine the critical field for Kondo peak splitting. While the onset for splitting in \0t (curvature crossing through zero) is already observed for the lowest field applied, for \1t and  \2t the critical field is $2\,\mathrm{T}$ and $4\,\mathrm{T}$, respectively. Similarly, the hydrogenated fragment (our canonical $S\!=\!1/2$ single $\pi$-radical) Kondo peak starts to split at $4\,\mathrm{T}$. This behavior shows that their Kondo state is compatible with a $\mathrm{S}\!=\!1/2$ fully Kondo screened ground state and corroborates that, in contrast to \0t, \1t and \2t lie in a neutral ground state on the Au(111) surface.

\clearpage





\subsection{Comparison between $I$, $\dd I/\dd V$, and CITS Kondo maps for \2t}

\begin{figure}[th!]
        \centering
    	\includegraphics[width=\textwidth]{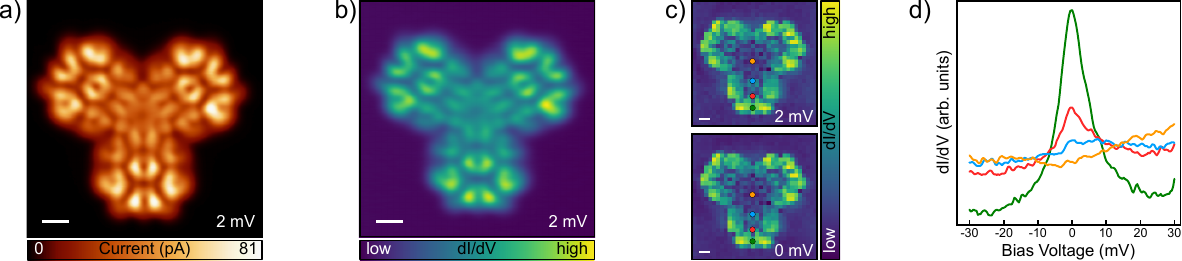}
        \caption{Constant-height a) current and b) conductance Kondo maps of \2t. c) Selected $\dd I/\dd V$ maps at 2 mV and 0 mV extracted from the CITS of \2t. All maps were acquired with a CO tip. Scalebars: 4 \si{\angstrom}. \rev{d) Point spectra extracted from the CITS over the positions marked by the colored dots in c)}.}
        \label{fig:FigS_KondoMap}
 \end{figure}

In Fig.~3b of the main manuscript, we experimentally map the Kondo orbitals by acquiring a constant-height CO image of the molecules at a very low bias voltage and tunnel junction resistances of approximately 25 M$\Omega$. In STM, molecular orbitals are typically visualized through a lock-in amplifier's $\dd I/\dd V$ demodulated signal. However, when the DC bias voltage is comparable to the amplitude of the AC signal, both the current and the $\dd I/\dd V$ demodulated signal produce the same image, as shown in Fig.~\ref{fig:FigS_KondoMap}a and b for the case of \2t.
Another approach to visualizing molecular orbitals is current imaging tunneling spectroscopy (CITS), which involves measuring a matrix of $\dd I/\dd V$ point spectra over the molecule at constant height, always starting from the same tunneling setpoint. This method allows mapping the conductance evolution across the bias range covered by the point spectra, including at 0 bias voltage. In Fig.~\ref{fig:FigS_KondoMap}c, we present $\dd I/\dd V$ maps at 2 mV and 0 mV for \2t, which qualitatively reveal the same molecular distribution as in the previous maps.
Current Kondo maps, as shown in Fig.~\ref{fig:FigS_KondoMap}a, simplify the process by avoiding lock-in demodulation, resulting in faster acquisition and better image quality.
\rev{Individual spectra are plotted in Fig.~\ref{fig:FigS_KondoMap}d, showing how the zero-bias peak vanishes from the edge of the molecule towards the N center, confirming the localization of the Kondo state over the zigzag edges.}

\clearpage

\subsection{IETS mapping of the spin excitation of \1t}

We measured the derivative of the conductance \rev{from the second-harmonic-demodulated lock-in signal} to map the molecular distribution of the spin excitation of \1t. The resulting maps match the simulations using the natural transition orbitals (NTO) from the two doublet states to the excited quartet state. \rev{Minor deviations from ideal $D_{3h}$ symmetry in some experimental maps are attributed to tip-induced anisotropy in constant-current imaging and local adsorption variations on the herringbone-reconstructed Au(111) surface. No systematic symmetry lowering is observed, indicating that these effects are not intrinsic to the molecular electronic structure.}

 \begin{figure}[th!]
        \centering
    	\includegraphics[width=\textwidth]{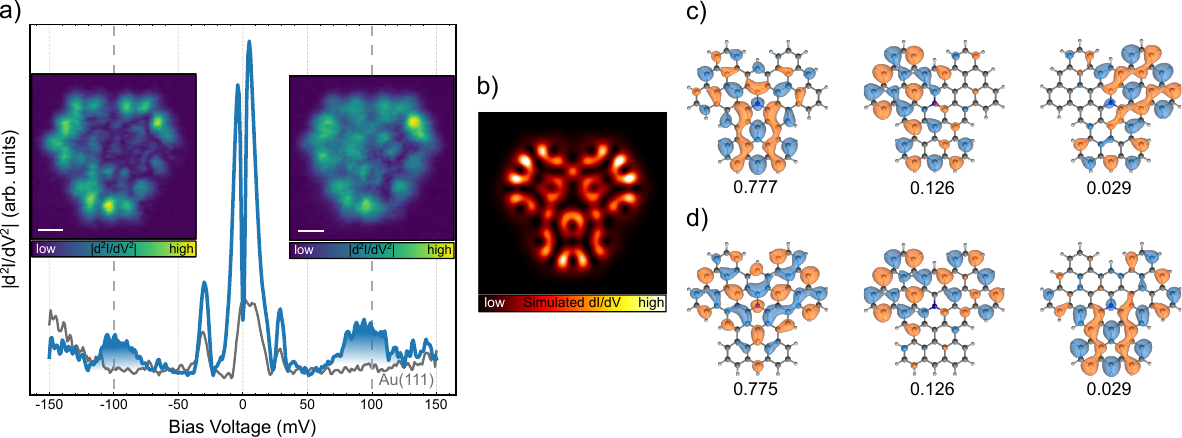}
        \caption{a) \rev{Absolute value of the derivative of the conductance measured over a zigzag edge of a \1t and on the bare surface. Modulation of $5\,\mathrm{mV}\,\mathrm{r.m.s}$. Insets: Experimental spin excitation maps at $\pm$100 mV measured at constant height. Scalebars: 4 \si{\angstrom}}. b) Simulated $\dd I/\dd V$ map obtained from the normalized weighted sum of the maps associated with each NTOs. c, d) Corresponding NTOs with their eigenvalues from the first and second doublet state to the quartet state, respectively, computed with \rev{CASSCF(11,11) wavefunction}.}
        \label{fig:FigS_spinexc}
 \end{figure}

 \clearpage

\subsection{Orbital mapping and Dyson orbitals}

\begin{figure}[th!]
        \centering
    	\includegraphics[width=\textwidth]{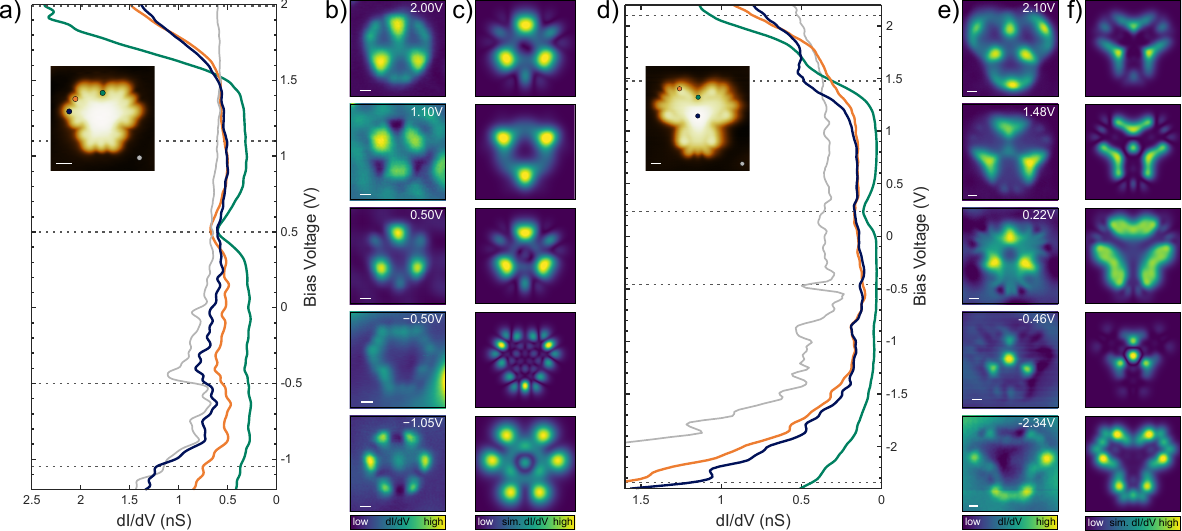}
        \caption{Long range STS spectra of a) \1t and d) \2t measured over several spots over the molecule and on the surface as marked by the colored dots in the STM image of the inset. \rev{Horizontal dashed lines indicate the bias values where the d$I$/d$V$ maps were measured}. b,e) Constant-current $\dd I/\dd V$ maps of the frontier orbitals of \1t and  \2t, respectively, \rev{(Scalebars: 4 \si{\angstrom}) and c,f) their comparison with simulated constant-height} $\dd I/\dd V$ maps of \1t and \2t obtained from the \rev{CASSCF} Dyson orbitals shown below \cite{KUMAR_Multireference_2025}.}     
        \label{fig:FigS_orbitals}
 \end{figure}

\rev{In Fig.~\ref{fig:FigS_orbitals}, we compare the spatial distribution of ionic resonances of \1t and \2t (Fig.~\ref{fig:FigS_orbitals}b and Fig.~\ref{fig:FigS_orbitals}e, respectively) with their computed Dyson orbitals shown in Fig.~\ref{fig:FigS_orbitals}c and Fig.~\ref{fig:FigS_orbitals}f. Strong agreement between theory and experiment is observed. This correspondence supports the assignment of a neutral charge state for both molecules on the surface.  Minor discrepancies between maps obtained at higher bias arise from the comparison between constant-current experimental maps and constant-height simulations, as well as from spectral broadening and mixing of nearby unoccupied states, and do not affect the assignment of the neutral ground state. }

 \begin{figure}[th!]
        \centering
    	\includegraphics[width=\textwidth]{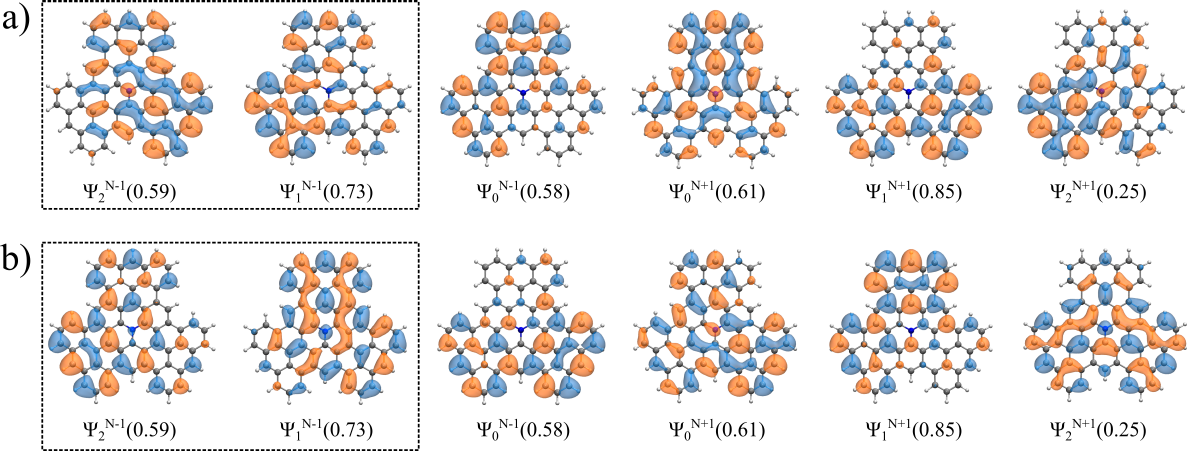}
        \caption{Dyson orbitals obtained from multireference \rev{CASSCF} calculations for the processes of single-electron addition and removal. The corresponding wavefunction norms are shown below for the \1t, using (a) the first doublet of the neutral state and (b) the second doublet of the neutral state. Orbitals highlighted with dashed boxes are degenerated. }
        \label{fig:dyson_aat}
 \end{figure}

 \begin{figure}[th!]
        \centering
    	\includegraphics[width=\textwidth]{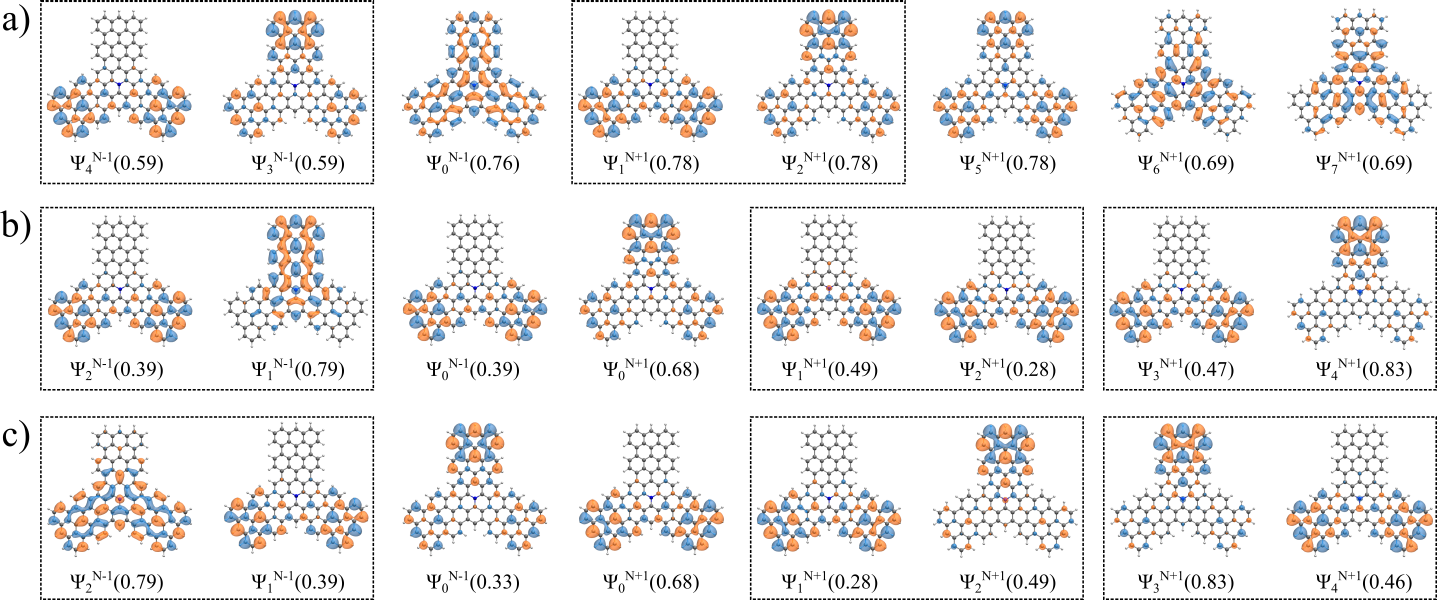}
        \caption{Dyson orbitals obtained from multireference \rev{CASSCF} calculations for the processes of single-electron addition and removal. The corresponding wavefunction norms are shown below for the \2t, using (a) the quartet of the neutral state, (b) the first doublet of the neutral state, and (c) the second doublet of the neutral state. Orbitals highlighted with dashed boxes are degenerated. }
        \label{fig:dyson_bat}
 \end{figure}

\clearpage



\subsection{Periodic DFT slab calculations and DMRG for 
Anderson model}
\begin{figure}[h!]
        \centering
    	\includegraphics[width=0.8\textwidth]{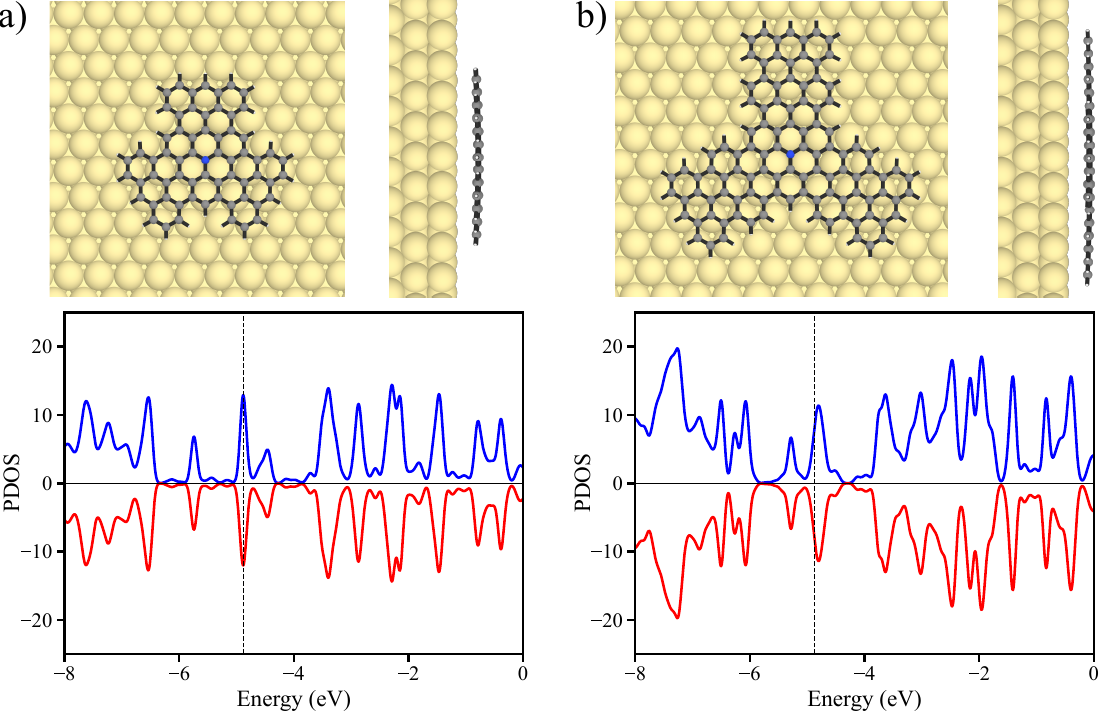}
        \caption{Structural configurations (top and side views) and molecular projected density of states (PDOS) for molecules adsorbed on a three-layer Au(111) surface: (a) \1t and (b) \2t. Blue and red lines represent the two spin components, shown with different signs for clarity. }
        \label{fig:FigS_SlabDFT}
 \end{figure}

From our measurements, both \1t and \2t molecules remain neutral on Au(111) with a spin doublet ground state. These observations contrast with the known triplet ground state of \0t on this surface, induced by the electron transfer into the substrate \cite{WANG_AzaTriangulene_2022}. As a starting point for understanding this charge state of \1t and \2t on Au(111), we have performed periodic slab DFT calculations (PBE+D3).  
These calculations (Fig.~\ref{fig:FigS_SlabDFT}) show that the single-occupied molecular orbitals (SOMOs), hosting three unpaired electrons in the free molecule, lie pinned to the Fermi level of Au(111) when adsorbed and lose an electron, becoming closed-shell. The convergence to a closed-shell electronic state on Au(111) is also obtained after imposing Broken-Symmetry initial guesses with different spin configurations at the beginning of the SCF cycle.  

The DFT results contradict our observations of spin-doublet ground states both \1t and \2t. The origin of this inconsistency is that the molecules under study have open-shell character and exhibit multireference wavefunctions, with many Slater determinants, which cannot be handled by DFT methods.  In fact, DFT-PBE tends to favor charge transfer when the singly occupied levels are not very far from the Fermi level due to the underestimation of the band gaps. Therefore, the use of multireference simulations in these systems is justified.

\begin{figure}[th!]
        \centering
    	\includegraphics[width=0.6\textwidth]{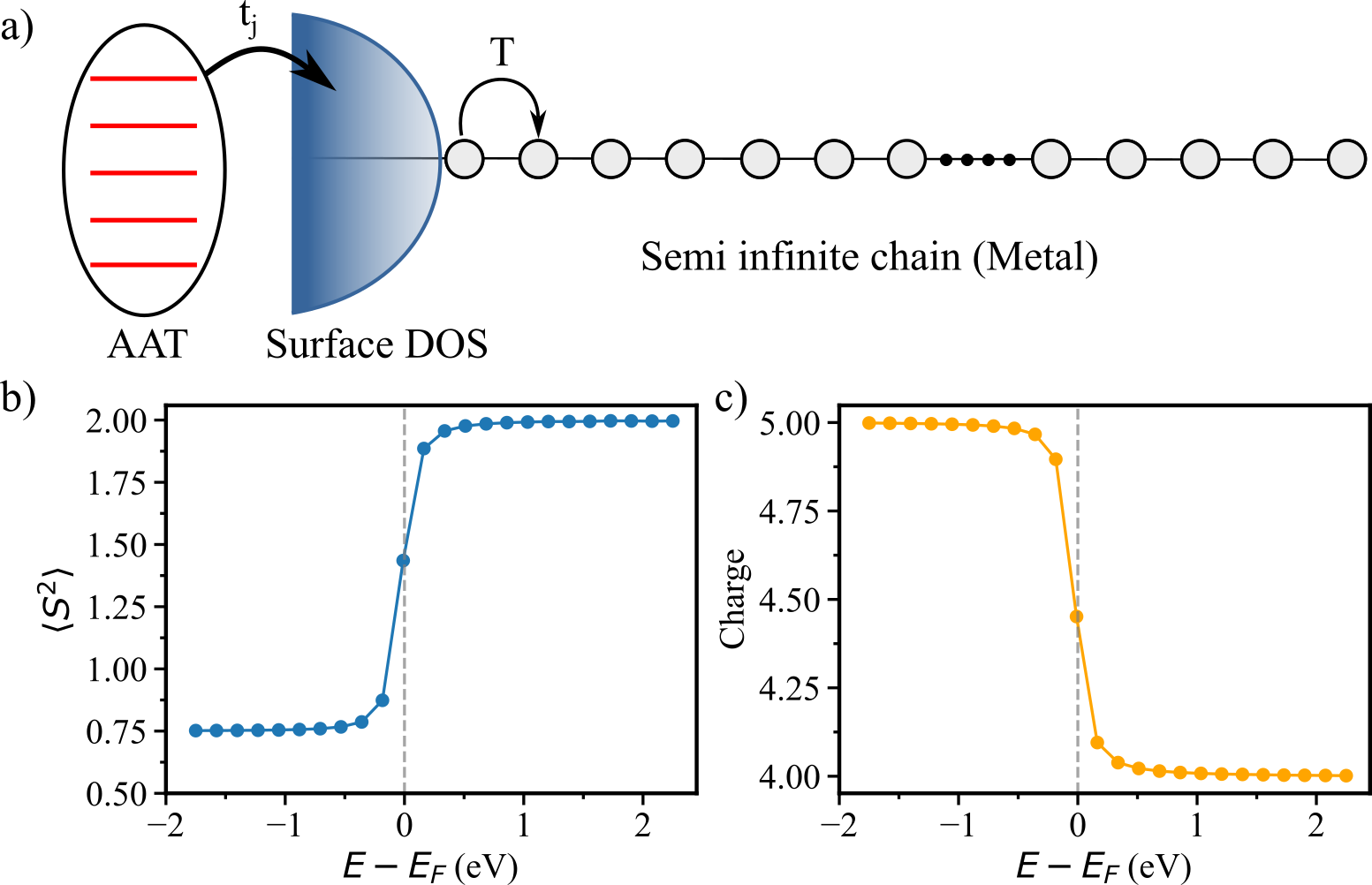}
        \caption{
        Schematic representation of the multichannel Anderson model used to couple the multireference molecular Hamiltonian of \1t to a 100-site metallic chain via density matrix renormalization group (DMRG), with coupling strength $t_j$. $T$ is the coupling between chain sites.  
        (b,c) Calculated $\langle S^2 \rangle$ and molecular charge as a function of energy relative to the Fermi level ($E_F$).
        }
        \label{fig:FigS_DMRG}
 \end{figure}

\newpage
\textbf{DMRG calculations to solve the Anderson model:} To understand the effect of charge transfer on \1t at the multireference level, we have solved the multichannel Anderson model using the density matrix renormalization group (DMRG), accounting for the metallic substrate within a many-body picture. We couple the full many-body molecular Hamiltonian obtained from \textit{ab initio} calculations to an effective chain consisting of 100 coupled sites that mimic the effect of the substrate (see Fig.~\ref{fig:FigS_DMRG}). These calculations explicitly account for charge fluctuations between the molecule and a metallic bath. The model reproduces that, when the SOMOs are shifted above the Fermi level, mimicking charge depletion, the system transitions to a triplet state $(S^2=2)$, consistent with the results of \0t, rather than to a closed-shell solution. In contrast, when the level remains below the Fermi energy, the molecule retains a multireference doublet ground state $(S^2=0.75)$. This clearly highlights that the spin multiplicity and ground state are governed by many-body effects that are not fully captured within a single-determinant framework, and the close-shell predicted by DFT-PBE is not the ground state of the molecule.

 \clearpage

\subsection{CASSCF calculations and Natural Orbitals}
\begin{figure}[th!]
\centering
\includegraphics[width=\textwidth]{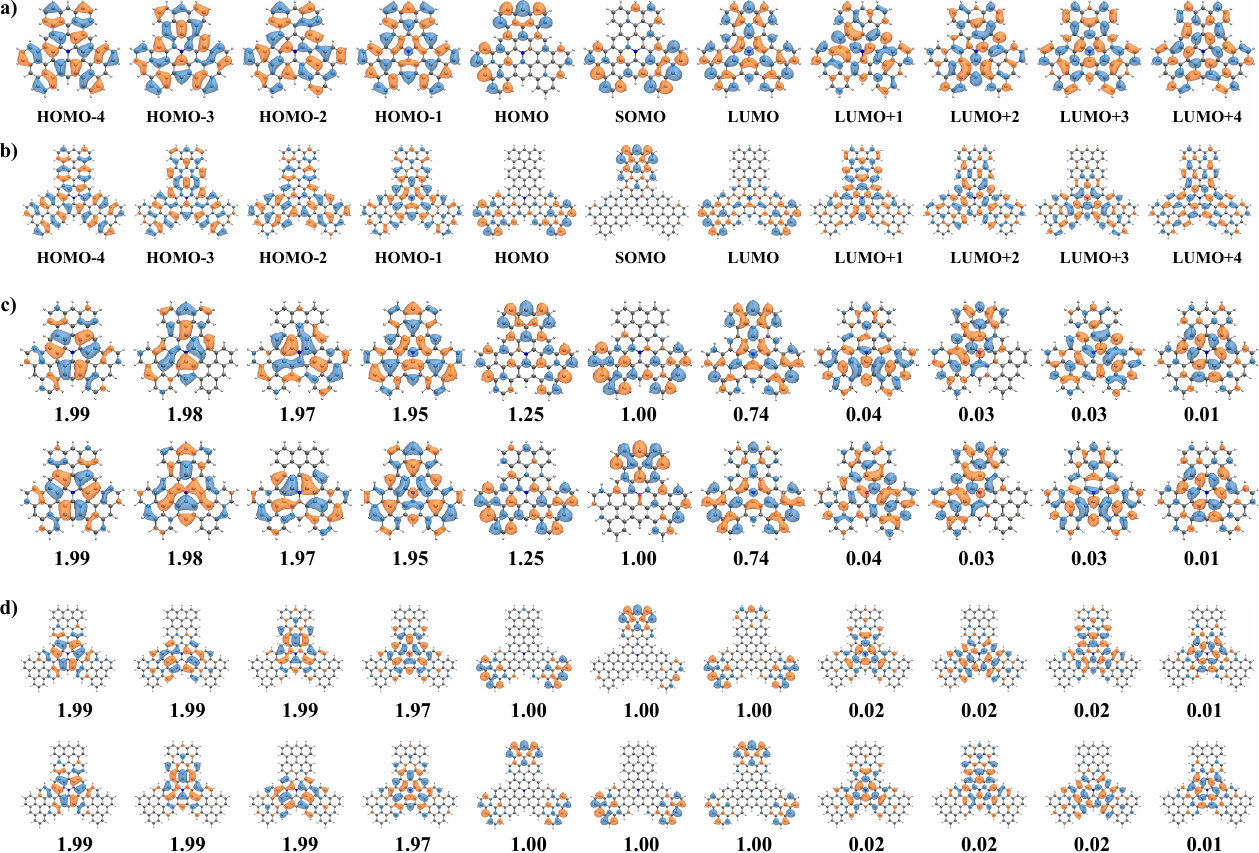}
\caption{(a,b) Restricted open-shell Kohn--Sham (ROKS) orbitals included in the CAS(11,11) active space for \1t and \2t, respectively. (c,d) Corresponding CASSCF NOs of the ground state for the first and second degenerate doublets.}
\label{fig:CASCI_natural_orb}
\end{figure}

\rev{
Complete Active Space Self-Consistent Field (CASSCF) calculations were performed using an active space of 11 electrons distributed over 11 orbitals, CAS(11,11), comprising the frontier molecular orbitals around the Fermi level. The restricted open-shell Kohn--Sham (ROKS) orbitals selected for the active space are shown in Fig.~\ref{fig:CASCI_natural_orb}(a,b) for \1t and \2t, respectively. The corresponding CASSCF natural orbitals (NOs) obtained for the ground-state wavefunction are presented in Fig.~\ref{fig:CASCI_natural_orb}(c,d).
}

\rev{
\textbf{Active-space convergence.} To assess the adequacy of the chosen active space, we examined the dependence of the doublet--quartet excitation energy and the NO occupations on the active-space size. For \1t, the excitation energy converges at the CAS(9,9) level, reaching approximately $120,\mathrm{meV}$, in excellent agreement with the experimentally observed IETS excitation ($\sim100,\mathrm{meV}$). Consistent with this energetic convergence, orbitals outside the active space acquire occupations close to 2 or 0, whereas the three frontier orbitals retain stable fractional occupations, indicating that the essential static correlation is fully captured.
}

\rev{
To further validate the multireference description, an active-space sensitivity analysis was performed for \0t, \1t, and \2t. Active spaces ranging from CAS(5,6) to CAS(9,9) and CAS(11,11) were considered, and both the NO occupations and the NEVPT2 spin-state energy gaps ($\Delta E_{\mathrm{d-d}}$ and $\Delta E_{\mathrm{d-q}}$) were monitored (Table~\ref{tab:active_space}). For all three molecules, the occupations of the strongly correlated orbitals (highlighted in bold) remain essentially unchanged when the active space is enlarged beyond CAS(5,6), indicating a stable multiconfigurational description.
}

\rev{
The NEVPT2 spin-state energy gaps converge at the CAS(9,9) level. In particular, for \1t, the calculated $\Delta E_{\mathrm{d-q}}$ increases from 86~meV in CAS(5,6) to 120~meV in CAS(9,9), with only a marginal change to 122~meV for CAS(11,11). Similar convergence is observed for \0t and \2t. These results demonstrate that CAS(9,9) is already sufficient to describe the relevant static correlation, while the larger CAS(11,11) employed throughout this work provides an additional safety margin and confirms that the calculated spin-state energetics and NO character are insensitive to the choice of active space.
}

\begin{table}[htpb]
\centering
\resizebox{\textwidth}{!}{%
\begin{tabular}{cccc}
\toprule
\textbf{Molecule} & \textbf{Active-space} & $\mathbf{\Delta E_{\text{d-d, d-q}}^{\text{(NEVPT2)}}}$ & \textbf{Occupation Natural Orbitals} \\
\midrule
\multirow{3}{*}{\0t}  & CAS(11,11) & 89, 782 & 1.97 1.97 1.95 1.94 \textbf{1.82 1.01 0.15} 0.07 0.06 0.05 0.02 \\
                     & CAS(9,9)   & 89, 782 & \phantom{1.97 }1.97 1.95 1.94 \textbf{1.82 1.01 0.15} 0.07 0.06 0.05 \\
                     & CAS(5,6)   & 89, 781 & \phantom{1.97 1.97 1.95 }1.94 \textbf{1.82 1.01 0.15} 0.07 0.06 \\
\midrule
\multirow{3}{*}{\1t} & CAS(11,11) & 0, 122 & 1.99 1.98 1.97 1.95 \textbf{1.25 1.00 0.74} 0.04 0.03 0.03 0.01 \\
                     & CAS(9,9)   & 0, 120 & \phantom{1.99 }1.98 1.98 1.95 \textbf{1.25 1.00 0.74} 0.04 0.03 0.02 \\
                     & CAS(5,6)   & 0, 86  & \phantom{1.99 1.98 1.98 }1.98 \textbf{1.27 1.00 0.73} 0.02 0.01 \\
\midrule
\multirow{3}{*}{\2t} & CAS(11,11) & 0, 0   & 1.99 1.98 1.98 1.96 \textbf{1.00 1.00 1.00} 0.04 0.03 0.03 0.01 \\
                     & CAS(9,9)   & 0, 0   & \phantom{1.99 }1.98 1.98 1.97 \textbf{1.00 1.00 1.00} 0.03 0.02 0.02 \\
                     & CAS(5,6)   & 0, 0   & \phantom{1.99 1.98 1.98 }1.98 \textbf{1.00 1.00 1.00} 0.01 0.01 \\
\bottomrule
\end{tabular}%
}
\caption{Active-space sensitivity analysis. NEVPT2 energy differences ($\Delta E_{\text{D-D, D-Q}}$) and CASSCF natural orbital occupations for molecules \0t, \1t, and \2t. The most strongly correlated natural orbital occupations are emphasized in bold.}
\label{tab:active_space}
\end{table}

\subsection{Kondo orbitals simulations}
The spatial distribution of the Kondo resonance was theoretically investigated through Kondo orbital simulations based on a multi-channel Anderson model. The Hamiltonian was diagonalized considering the many-body multiplet structure derived from \rev{CASSCF} calculations for the neutral ground state and virtual charge states, following the methodology in Ref.~\cite{CALVO-FERNANDEZ_Theoretical_2024}. Virtual processes involving the five lowest-energy multiplets in charged states were included to compute the scattering amplitudes.

\begin{figure}[th!]
        \centering
    	\includegraphics[width=\textwidth]{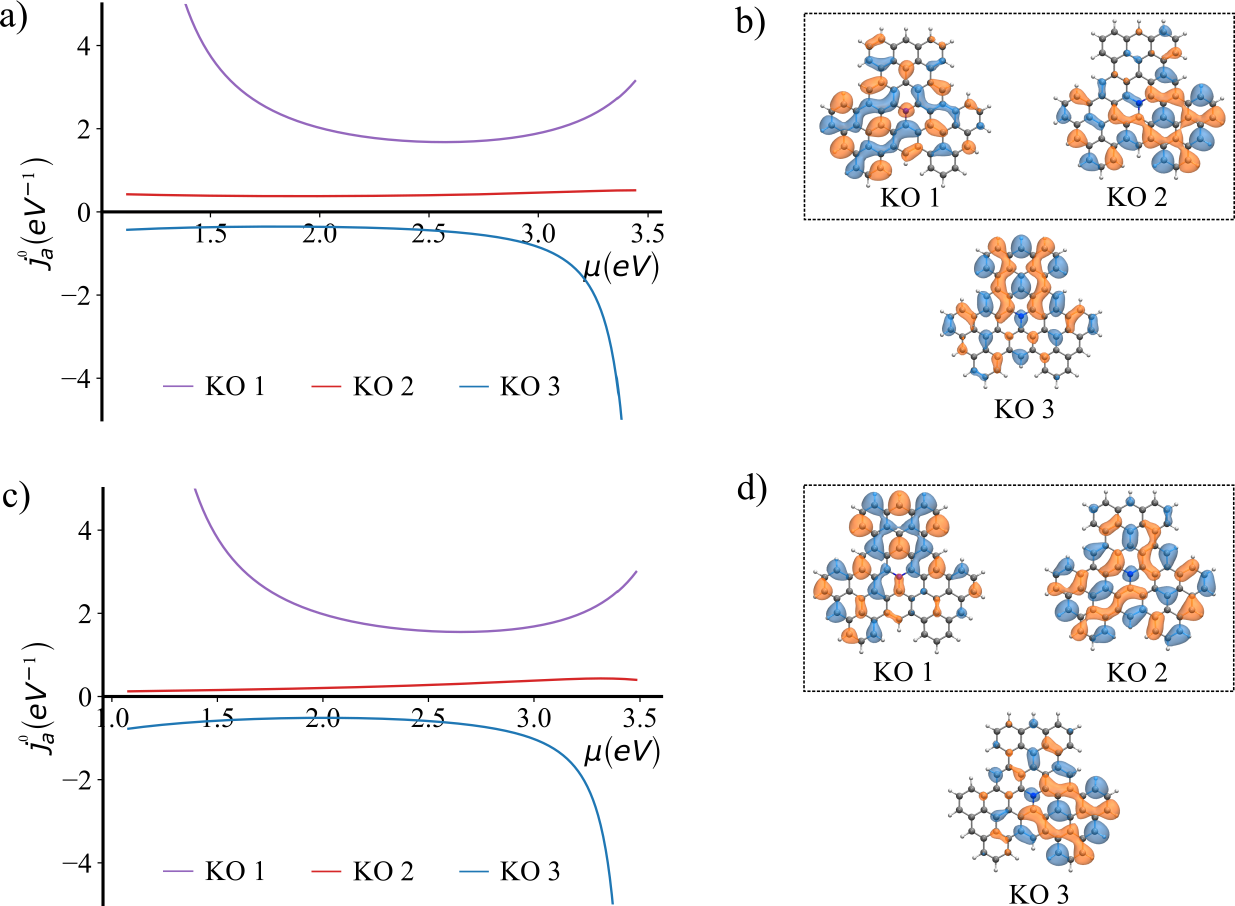}
        \caption{Results from the multi-orbital Kondo analysis used to simulate the spatial Kondo distribution map presented in the main text.
(a) Coupling constants $j_a$ as a function of chemical potential for each Kondo orbital (KO). (b) Orbital isosurfaces of the KOs that exhibit non-zero coupling to the substrate conduction electrons at the point of electron–hole symmetry. Orbitals highlighted with dashed boxes denote channels with antiferromagnetic coupling ($j_a > 0$) to the electron bath, and thus participate in the many-body Kondo screening process for the first doublet state.
(c,d) Corresponding results for the second doublet state for \1t.}
        \label{fig:CASCI_kondo_orb}
 \end{figure}

For \1t, the analysis of coupling constants versus chemical potential (Fig.~\ref{fig:CASCI_kondo_orb}a) identifies specific orbital channels with antiferromagnetic coupling ($j_a>0$) that participate in the Kondo screening. The corresponding orbital isosurfaces at electron-hole symmetry (Fig.~\ref{fig:CASCI_kondo_orb}b) reveal the spatial patterns of these active channels for the first doublet state. Similar analysis for the second doublet state (Fig.~\ref{fig:CASCI_kondo_orb}c,d) shows distinct orbital participation in the screening process. As the ground state is doubly degenerated, consequently, the final Kondo spatial maps are constructed as a combination of the antiferromagnetic channels from both doublets.

\begin{figure}[th!]
        \centering
    	\includegraphics[width=\textwidth]{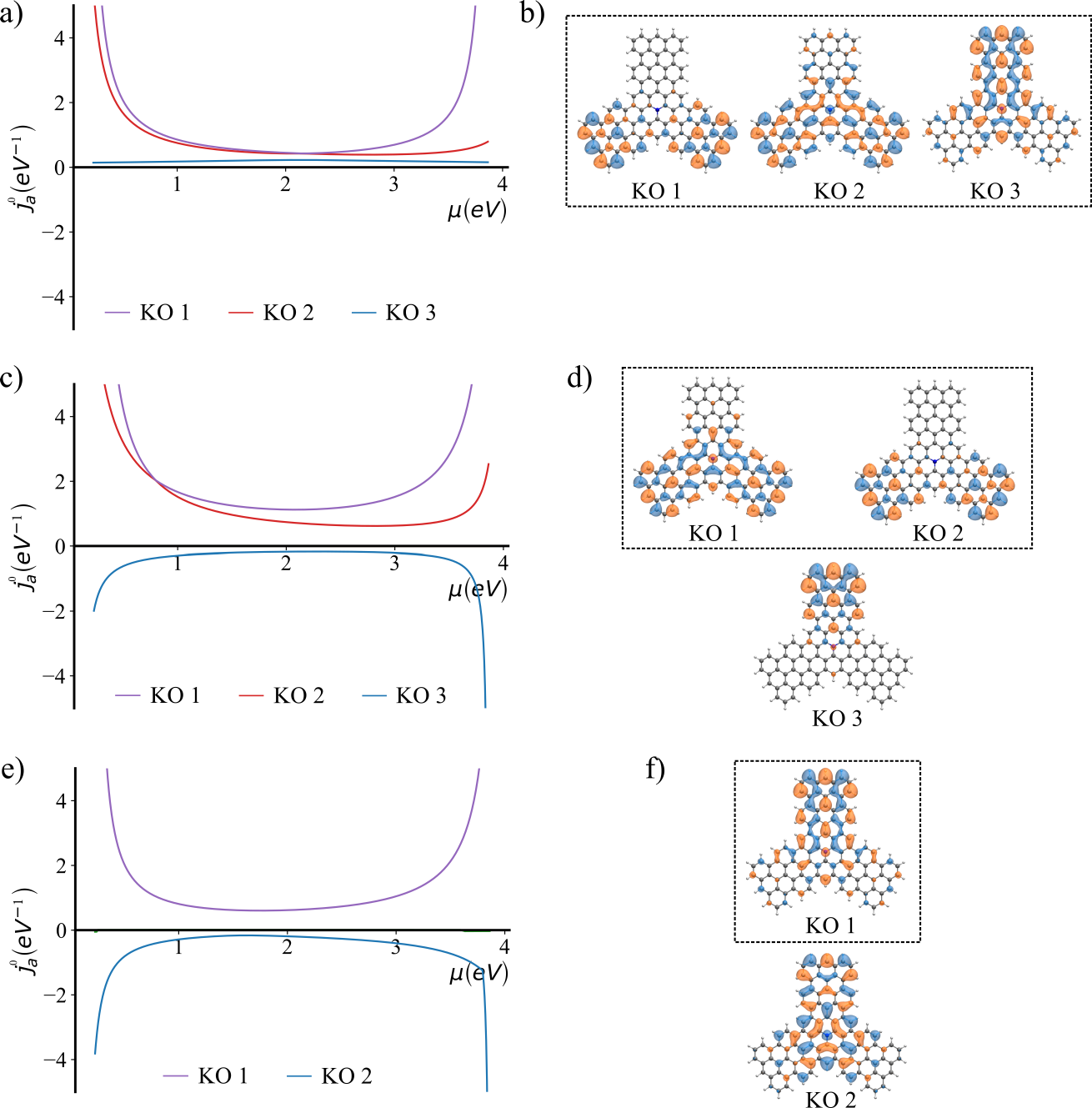}
        \caption{Results from the multi-orbital Kondo analysis used to simulate the spatial Kondo distribution map presented in the main text.
(a) Coupling constants $j_a$ as a function of chemical potential for each Kondo orbital (KO). (b) Orbital isosurfaces of the KOs that exhibit non-zero coupling to the substrate conduction electrons at the point of electron–hole symmetry. Orbitals highlighted with dashed boxes denote channels with antiferromagnetic coupling ($j_a > 0$) to the electron bath, and thus participate in the many-body Kondo screening process for the quartet state.
(c,d) Corresponding results for the first doublet state, and (e,f) for the second doublet state for \2t.}
        \label{fig:CASCI_kondo_BAT_orb}
 \end{figure}

Similarly, for \2t molecule, the Kondo analysis provides the three antiferromagnetic coupling channels ($j_a >0$) from the quartet state (Fig.~\ref{fig:CASCI_kondo_BAT_orb}a), while two antiferromagnetic channels from the first doublet and one antiferromagnetic channel from the second doublet. The CASSCF calculations reveal a near-degeneracy between the lowest-lying quartet state and two doublet states. This near-degeneracy implies that the effective low-energy manifold cannot be described by a single spin state, but instead arises from the interplay of the quartet and both doublets. As a result, the Kondo spatial distribution maps incorporate contributions from all three states.

\clearpage

\subsection{About the Jahn-Teller distortion and dynamic effects}

To elucidate the absence of the Jahn–Teller distortion in molecules \1t and \2t, we optimized the molecular geometries in both the doublet and quartet spin states using the PBE0 functional. The PBE0 calculations of \1t predict a distorted structure ($C_{2v}$) in the doublet state consistent with a Jahn–Teller effect, which is lower in energy by $332\,\mathrm{meV}$ relative to the symmetry ($D_{3h}$) geometry obtained in the quartet state. However, the PBE0 solution exhibits broken symmetry in the molecular orbitals, leading to substantial spin contamination. This is expected because a doublet state with three effectively unpaired electrons cannot be described adequately by a single Slater determinant. As a result, single-determinant DFT may provide an incomplete description of the electronic structure.

\begin{figure}[bh!]
        \centering
    	\includegraphics[width=0.4\textwidth]{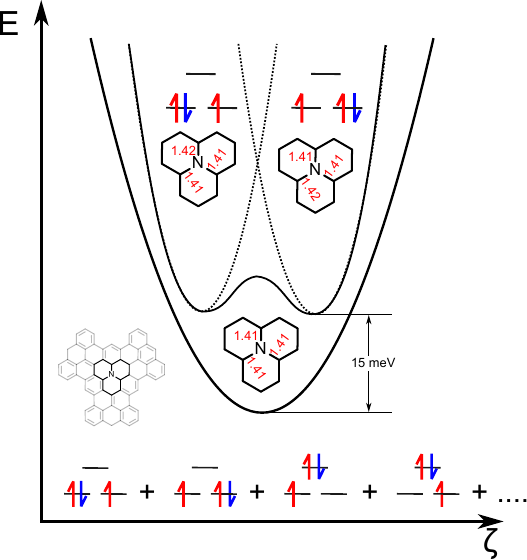}
        \caption{\textbf{Schematic potential energy surface and dynamic Jahn–Teller effect.} Calculated electronic energy $E$ as a function of the distortion coordinate $\zeta$ for the doublet ground state. Dashed lines indicate the potential energy predicted by single-determinant DFT (PBE0), which incorrectly favors a deep $C_{2v}$ distorted minimum characterized by asymmetric bond lengths (\SI{1.42}{\angstrom} vs.~\SI{1.41}{\angstrom}). The solid line represents the multireference CAS(11,11) solution, which captures the multi-configurational nature of the unpaired electrons. The resulting shallow potential energy surface (with an energy difference of only $\sim\!\SI{15}{meV}$) is indicative of a dynamic Jahn-Teller regime, where rapid structural fluctuations lead to an experimentally observed time-averaged $D_{3h}$ symmetry.}
        \label{fig:FigS_JT}
 \end{figure}

To address this limitation, we performed multireference calculations for both non-symmetric ($C_{2v}$) and symmetric ($D_{3h}$) geometries. Within a complete active space framework, CAS(11,11), the electronic wavefunction is expanded over all Slater determinants, capturing the strong multireference character of the electronic structure associated with the three unpaired electrons. In this multireference treatment, the symmetric structure ($D_{3h}$) is found to be $15\,\mathrm{meV}$ lower in energy than the distorted geometry ($C_{2v}$). 
It is crucial to note that this $15\,\mathrm{meV}$ energy difference is small and falls well within the intrinsic error margin of the CASSCF/NEVPT2 methods. Therefore, a strictly rigid $D_{3h}$ symmetric ground state cannot be definitively concluded from these calculations alone. Instead, this near-degeneracy strongly suggests that the system may as well reside on a shallow, nearly barrierless ``Mexican hat" potential energy surface, characteristic of a dynamic Jahn–Teller (JT) effect (see Fig.~\ref{fig:FigS_JT}). In the experiment, the energy resolution is set by the width of the Kondo resonance, which is about 1 meV. Consequently, the scenario of the dynamic Jahn-Teller effect is possible, where the system undergoes rapid quantum or thermal pseudorotation between equivalent distortions. Spatial probes like our STS and Kondo maps inherently time-average this motion, projecting an apparent D3h symmetry even if the instantaneous structure is distorted.

 \clearpage



\subsection{Calculated vibrational energies and IR absorption spectra and electron-phonon coupling}

\begin{figure}[th!]
        \centering
    	\includegraphics[width=1\textwidth]{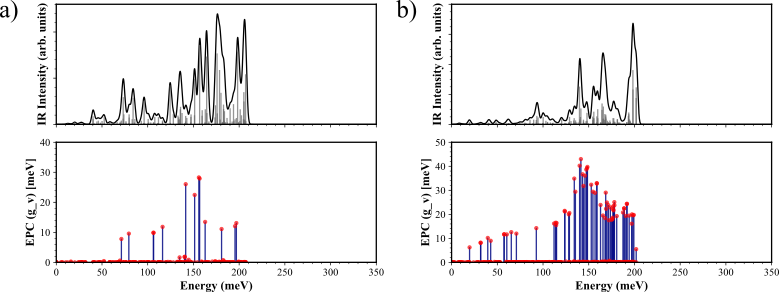}
        \caption{ \textbf{Infrared Spectra and Electron–Phonon Coupling in \1t and \2t.} Comparative analysis of vibrational modes and their corresponding electron–phonon coupling strengths for AAT (a, left) and BAT (b, right). The top panels show simulated infrared spectra as a function of energy, with gray vertical bars representing discrete vibrational transition intensities and the solid black curves denoting the broadened spectral envelopes. The bottom panels present the corresponding electron–phonon coupling constants ($g_\nu$) for each vibrational mode, with coupling strengths indicated by red markers on blue drop-lines.
}
        \label{fig:FigS_IR_SPECTRUM}
 \end{figure}

To provide further support for the magnetic origin of the $\sim\!100$~meV signal in \1t, we compare the structural and vibrational relationship between \1t and its extended homologue, \2t. Because \2t is a direct structural extension of \1t, and both molecules share fundamentally similar carbon frameworks and bonding motifs. Consequently, their vibrational normal modes are expected to be highly comparable in this energy regime. This is supported by our calculated IR absorption spectra (Fig.~\ref{fig:FigS_IR_SPECTRUM}), which demonstrate analogous vibrational distributions for both molecules, albeit with a higher density of modes for \2t due to its larger size. 
To further evaluate the potential vibrational contribution to the experimentally observed IETS features, we computed the electron-phonon coupling (EPC) constants ($g_\nu$) for each normal mode of both molecules. These calculations provide a quantitative measure of the coupling strength between electronic states and specific molecular vibrations. The EPC constants were calculated using a finite-difference approach based on vibrational frequencies and Cartesian normal mode displacement vectors obtained from gas-phase calculations in ORCA.

For each vibrational mode $\nu$ with angular frequency $\omega_\nu$, the optimized molecular geometry was displaced along the corresponding Cartesian normal mode vector $v_{i,\nu}$ by a fixed amplitude of $\pm\delta q = \pm0.01$~\AA. The mass-weighted normal coordinate displacement, $\Delta Q_\nu$, was computed as:
\begin{equation}
\Delta Q_\nu = \sqrt{\sum_i m_i \Delta x_{i,\nu}^2}
\end{equation}
where $m_i$ is the atomic mass of atom $i$, and $\Delta x_{i,\nu} = \delta q \cdot v_{i,\nu}$ is the Cartesian displacement. Multireference CASCI(11,11) energy calculations were then performed at these displaced geometries to obtain the energies $E(+\Delta Q_\nu)$ and $E(-\Delta Q_\nu)$. The linear deformation potential was evaluated using the central finite-difference method~\cite{coropceanu2007charge, devos1998electron}:
\begin{equation}
\frac{\partial E}{\partial Q_\nu} \approx \frac{E(+\Delta Q_\nu) - E(-\Delta Q_\nu)}{2 \Delta Q_\nu}
\end{equation}
The electron-phonon coupling constants $g_\nu$ were calculated by multiplying the deformation potential by the zero-point vibrational amplitude:
\begin{equation}
g_\nu = \frac{\partial E}{\partial Q_\nu} \sqrt{\frac{\hbar}{2\omega_\nu}}
\end{equation}

As shown in the bottom panels of Fig.~\ref{fig:FigS_IR_SPECTRUM}, the calculated EPC strengths remain modest across the entire vibrational spectrum for both molecules, with no exceptional coupling strength observed near 100~meV. This further supports that vibrational excitations are unlikely to dominate the IETS signal in this energy range.

Critically, if the experimentally observed $\sim\!100$~meV feature in \1t were merely a structural vibrational mode, a structurally analogous molecule like \2t should inevitably exhibit a corresponding inelastic step at the same energy. However, this distinct $\sim\!100$~meV signature is only observed in \1t. This definitive presence in \1t and notable absence in \2t  argue against a purely vibrational origin. Instead, it provides compelling evidence that the $\sim\!100$~meV signal corresponds to a spin excitation arising from frustrated spin interactions specific to the \1t electronic structure, as also shown from the many-body CASSCF calculation, thereby distinguishing its magnetic ground state from the quasi-independent edge spins characteristic of \2t.

\newpage
\subsection{Hubbard toy-model}
\label{sec:toymodel}

\rev{To rationalize the antiparallel magnetic exchange pattern in \1t,  we use a four-site Hubbard model (Fig. 4d in the manuscript). This \textit{toy-model} represents the effective minimal description of the low-energy spin manifold of triangular molecules \cite{MISHRA_Observation_2021}. In the platforms studied here, each site ideally represents a topological zero-energy state (ZES), three of them at the termini of the 7AGNR extensions \cite{CAO_Topological_2017}, and the fourth one, at the junction among them \cite{TAMAKI_Topological_2020}. The central site is essential to capture the electronic configuration of a five-electron correlated manifold similar to CAS, with the central state mediating the type of interaction between the three edge radicals. }

\begin{figure}[bh!]
        \centering
    	\includegraphics[width=\textwidth]{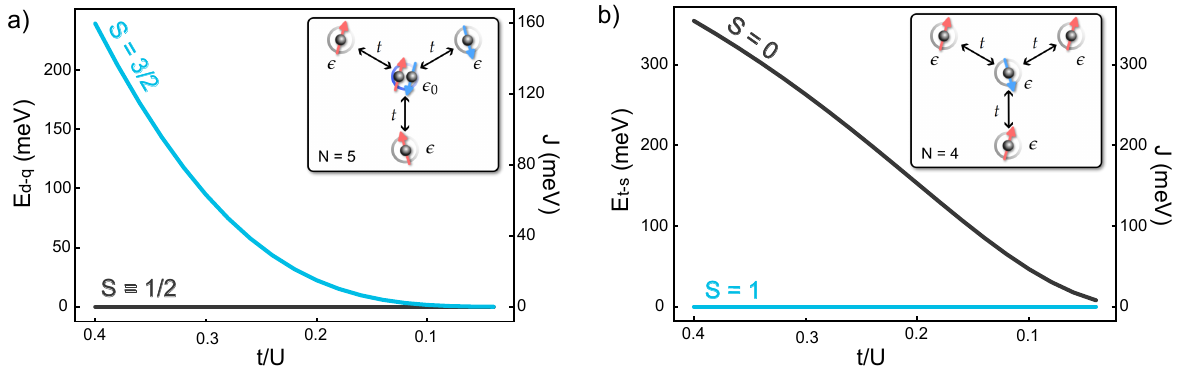}
        \caption{Difference in energy between the excited and ground states \rev{and the corresponding magnetic exchange parameter ($E_{d\text{-}q}=3/2J$~\cite{HARALDSEN_Neutron_2005} or $E_{t\text{-}s}=J$)} obtained from the four-site Hubbard toy model as a function of hopping parameter $t$ for the case of (a) $\epsilon_0 /U = -1.2$ and $N=5$, simulating an aza-core with an extra electron provided by the N-doping at the center, and (b) $\epsilon_0/U=0$ and $N=4$, simulating the equivalent case with a carbon atom at the center (undoped).  
        The exchange pattern is completely reversed by the N substitution. Schematic representations of each model with their approximate occupations are shown as insets.}
        \label{fig:FigS_Hubbard}
 \end{figure}

Insets in Fig.~\ref{fig:FigS_Hubbard} schematically show the model. The Hamiltonian reads

\begin{equation*}
\mathscr{H} = -t \sum_{i=1,\sigma}^3 (c_{i,\sigma}^\dagger c_{0,\sigma} + \mathrm{H.c.})
+ U \sum_{i=0}^3 n_{i\uparrow} n_{i\downarrow}
+ \sum_{i=0,\sigma}^3 \epsilon_i n_{i,\sigma},
\end{equation*}
where the first term corresponds to electron hopping with amplitude $t$, the second to on-site Coulomb repulsion, and the third to the on-site energies $\epsilon_i$, taken as zero \rev{for the singly occupied ZES.  To simulate the aza-core in \1t, we use a negative on-site energy $\epsilon_0$ for the central site,}  and consider only the subspace with five electrons ($N=5$), while, for the all-carbon analogue, all on-site energies are set equal to $\epsilon_0=0$ and $N=4$. 

\begin{figure}
    \centering
    \includegraphics[width=0.45\textwidth]{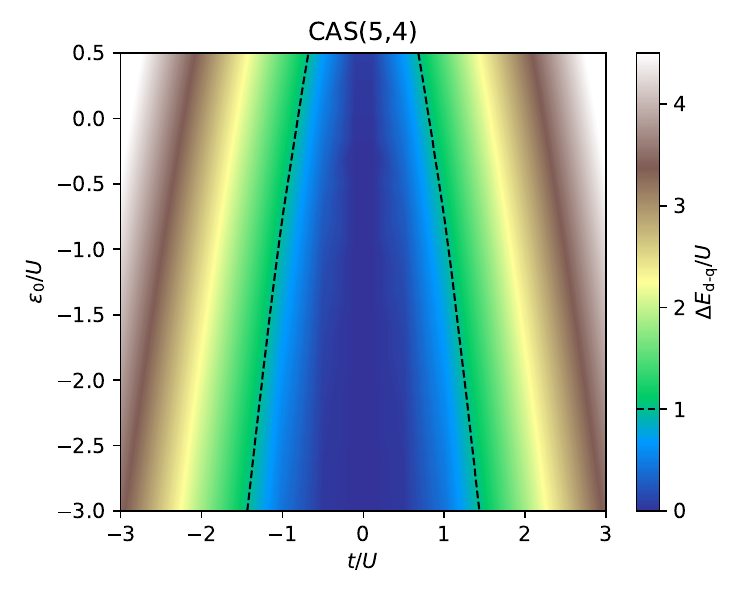}
    \includegraphics[width=0.45\textwidth]{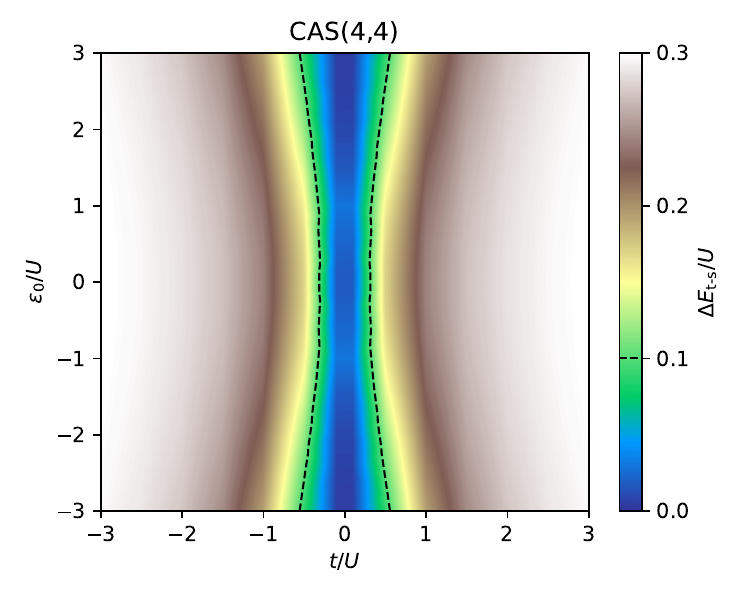}
    \caption{\rev{Excitation energy $\Delta E_\text{d-q}$ (left panel) and $\Delta E_\text{t-s}$ (right panel) within the four-orbital toy model using exact diagonalization, considering different model parameters (effective hopping $t$ and centre orbital energy $\epsilon_0$ in units of the Coulomb repulsion $U$) and electron filling $N_\text{elec}$.
    CAS(5,4) thus corresponds to the model presented in the main text for the nitrogen-containing molecules exhibiting a doublet ground state, while 
    CAS(4,4) would be the corresponding model for an all-carbon analog (one electron less) with a triplet ground state.
    The dashed contour lines are guides to the eye for $\Delta E_\text{d-q}=U$ and $\Delta E_\text{t-s}=0.1 U$, respectively.}}
    \label{fig:FigS_Hubbard_cas}
\end{figure}

The Hamiltonian was solved numerically by exact diagonalization, yielding the energies and total spin of the ground and excited states  \rev{ as a function of $t$, $U$ and $\epsilon_0$; we use $U$ as a reference energy scale. Figures~\ref{fig:FigS_Hubbard}a and \ref{fig:FigS_Hubbard}b show map the excitation energy as a function of $t/U$ for the N-doped ($N = 5, \epsilon_0 /U = -1.2$~eV) and all-carbon cases ($N = 4$, $\epsilon_0 = 0$), respectively.  Fig.~\ref{fig:FigS_Hubbard_cas} explores the effect of$\epsilon_0/U$. }

In the former, the ground state is a doubly degenerate doublet, and the excitation energy to the lowest quartet state decreases monotonically with $t/U$, which agrees with the CAS results. In this configuration, the three unpaired electrons at the edges form a frustrated spin trimer, with antiferromagnetic exchange mediated by \rev{the tendency of the aza-core to be doubly occupied}. This exchange pattern reversesin the all-carbon platform. \rev{In this case, the ground state is a high-spin triplet that fulfills Lieb's theorem with three ferromagnetically aligned terminal spins in the majority sublattice} and a singly occupied junction state~\cite{TAMAKI_Topological_2020}, antiferromagnetically coupled to them. The excitation energy to the two degenerate singlet states also decreases monotonically with the hopping parameter $t/U$. 

\rev{These results show the relevant effect of N substitution to chemically control intramolecular exchange patterns. Representing the aza core as a fifth electron added into the four-site Hubbard model and a reduced on-site energy at the central site reverses the exchange coupling between the terminal spins, producing AFM alignment despite their location on the same sublattice. This exchange pattern does not contradict Lieb's theorem, which no longer applies since the lattice is not half-filled \cite{LIEB_Two_1989}. }

\rev{\textbf{Mapping the correlated electron states into a Heisenberg model}: The results can further be mapped into an interacting spin system through a Heisenberg  Hamiltonian \cite{HARALDSEN_Neutron_2005}. The ground state of an AFM Heisenberg spin trimer is a degenerate doublet with doublet-quartet excitation energy $E_{d\text{-}q}=3/2J$ (Fig.~\ref{fig:FigS_Hubbard}), where $J$ represents the exchange interaction between edge spins. }

\clearpage

\newpage

\subsection{CAS-Hubbard calculations}
To complement the calculations in the previous section \ref{sec:toymodel}, we also consider an analysis of the excitation spectra based on the Hubbard Hamiltonian
for a full $\pi$-electron model:
\begin{equation}
	H = \sum_{\langle ij\rangle ,\sigma} h_{ij} c^\dagger_{i\sigma} c_{j\sigma} 
	  + U \sum_{i} n_{i\uparrow}n_{i\downarrow},
\end{equation}
where $n_{i\sigma}=c^\dagger_{i\sigma} c_{i\sigma}$ and $c_{i\sigma}$ are the electronic number and annihilation operators, respectively, corresponding to atomic site $i$ and spin $\sigma$.
The matrix $h_{ij}$ includes the onsite potentials $\varepsilon_i$ along the diagonal and hopping matrix elements $t_{ij}$ off the diagonal. $U$ denotes the onsite Coulomb repulsion for a carbon or nitrogen $2p_z$ orbital.
We include up to third-nearest neighbor hopping with $(t_1, t_2, t_3)=-(2.7,0.2,0.18)$ eV following a parametrization for graphene nanoribbons from Ref.~\cite{HaUpSa.10.Generalizedtightbinding}.
For the electrostatic potential associated with the nitrogen atom, we assume a Gaussian variation in the onsite energies $\varepsilon_i(\rho) = \varepsilon_N e^{-(\rho/R)^2}$ at a distance $\rho$ away from the central heteroatom. Here $\varepsilon_N=-5.5\,\mathrm{eV}$ and $R=1.5\,\mathrm{\AA}$ are \textit{effective} model parameters that characterize the depth and extension of the potential variation, respectively, with values chosen to provide the best fit to the excitation energies seen experimentally ($\sim100\,\mathrm{meV}$ for \1t and $\sim0\,\mathrm{meV}$ for \2t and for an even longer teranthene aza-triangulene (\3t)).

\begin{figure}[th!]
    \centering
    \includegraphics[width=\textwidth]{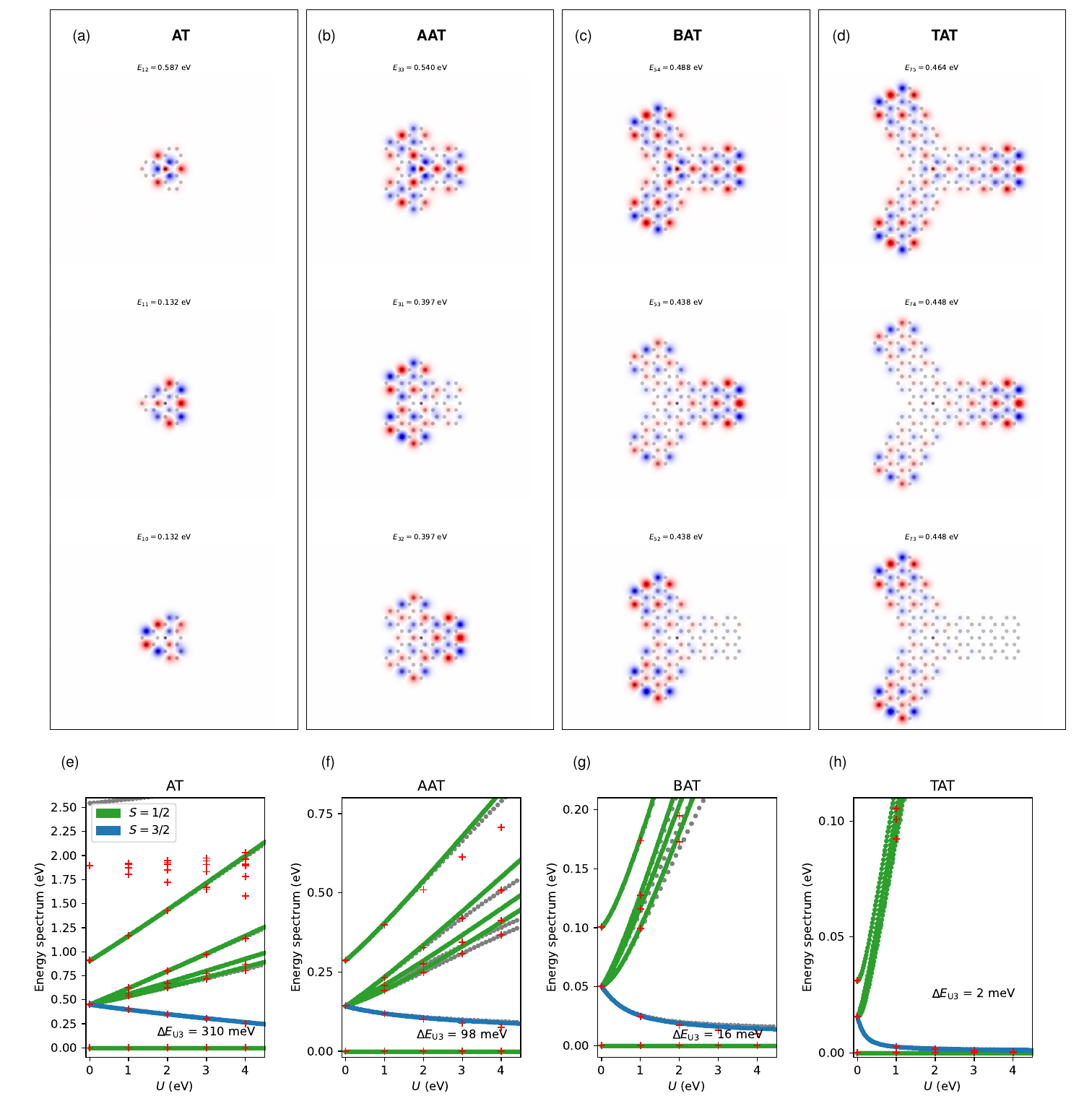}
    \caption{(a)-(d) Visualization of the active orbitals as included in CAS(3,3) with the Hubbard Hamiltonian. The sign and real-space intensity is shown with a red/blue colormap. The gray dots indicate the atomic positions. The first two orbitals (lower two rows) are degenerate and have no weight on the central nitrogen atom. In contrast, the third orbital (top row) is nondegenerate and has weight on nitrogen.
    (e)-(h) Calculated energy spectrum. The green/blue dots correspond to the minimal CAS(3,3) description. The gray dots [red crosses] correspond to more accurate CAS(5,4) [CAS(11,10)] calculations, essentially providing the same result for the low-energy states.
	The label $\Delta E_\text{U3}$ indicates the excitation gap at the CAS(3,3) level at $U=3\,\mathrm{eV}$.}
    \label{fig:CAS33}
\end{figure}

The Hubbard Hamiltonian may also be expressed in the eigenbasis $(\varepsilon_\alpha, \phi_\alpha)$ of $h_{ij}$ as
\begin{equation}
	H = \sum_{\alpha\sigma} \varepsilon_{\alpha} n_{\alpha \sigma} 
	+ U \sum_{\alpha\alpha'\beta\beta'} 
	\left(\sum_i \phi^*_{\alpha_i}\phi_{\alpha'_i}\phi^*_{\beta_i}\phi_{\beta'_i} \right) 
		c^\dagger_{\alpha\uparrow} c_{\alpha'\uparrow}c^\dagger_{\beta\downarrow} c_{\beta'\downarrow}.
\end{equation}
In this basis we explore the many-body energy spectrum with the complete active space (CAS) method, considering the active space CAS$(N_\text{elec}, N_\text{orb})$ for the number of included electrons and orbitals, respectively. The active orbitals are chosen around the singly-occupied molecular orbital (SOMO).
For describing the low-energy states of the present molecular systems based on the Hubbard Hamiltonian, we find it to be sufficient to consider just three electrons in the first three frontier orbitals, i.e., CAS(3,3) as shown in Fig.~\ref{fig:CAS33}. The accuracy of this minimal approach is bench-marked by increasing the active space to both CAS(5,4) and CAS(11,10), providing minimal changes to the excitation spectrum.

\begin{figure}[th!]
    \centering
    \includegraphics[width=\textwidth]{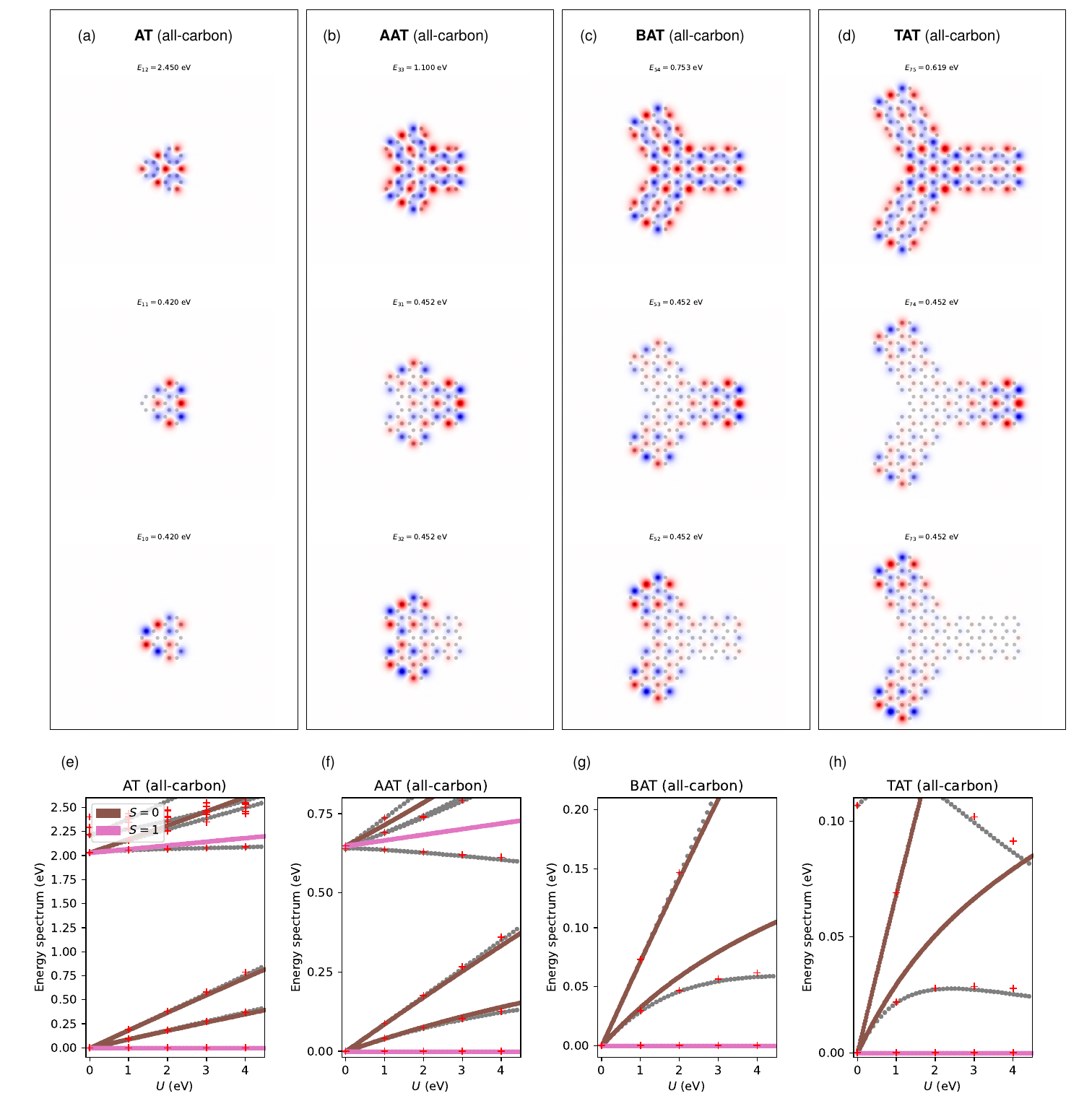}
    \caption{(a)-(d) Visualization of the active orbitals as included in CAS(2,3) with the Hubbard Hamiltonian for the all-carbon cases of Fig.~\ref{fig:CAS33}.
    (e)-(h) Calculated energy spectrum. The brown/pink dots correspond to the minimal CAS(2,3) description. The gray dots [red crosses] correspond to more accurate CAS(4,4) [CAS(10,10)] calculations.
    For \2t and \3t, CAS(2,3) significantly overestimates the first excitation gap.}
    \label{fig:CAS23}
\end{figure}

As shown in Fig.~\ref{fig:CAS23} we also explored the all-carbon analogues of the different molecules. Here the minimal model corresponds to CAS(2,3). However, for the larger molecules, it is necessary to include more orbitals in the active space for an accurate description of the first excitation gap. 

\clearpage

\bibliography{refs_SI}